\documentclass{article}
\usepackage{graphicx} 

\usepackage[backend=biber,style=numeric,sorting=none]{biblatex}
\usepackage{float}
\usepackage{amsmath, amssymb}
\usepackage[margin=2.5cm]{geometry}
\usepackage{hyperref}
\usepackage{authblk} 

\title{Learning Neural Operator Surrogates for the Black Hole Accretion Code}

\author[1]{Matthias Nägele\thanks{These authors contributed equally.}}
\author[2]{Cedric Bös\textsuperscript{*}}
\author[2]{Chester Tan}
\author[1]{Christian M. Fromm}
\author[2]{Ingo Scholtes}
\author[1]{Karl Mannheim}

\affil[1]{Institute for Theoretical Physics and Astrophysics, Julius-Maximilians-Universität Würzburg, Würzburg, Germany}
\affil[2]{Chair for Machine Learning for Complex Networks,
Julius-Maximilians-Universität
Würzburg, Würzburg, Germany}

\date{April 28$^{\text{th}}$, 2026}

\begin{document}

\maketitle

\begin{abstract}
General-relativistic magnetohydrodynamic (GR-MHD) simulations are essential for studying black hole accretion, relativistic jets, and magnetic reconnection, yet their computational cost severely limits systematic parameter exploration. We investigate neural operator surrogates for two astrophysically relevant simulation scenarios produced by the Black Hole Accretion Code (\texttt{BHAC}).

First, a Physics Informed Fourier Neural Operator (PINO) is trained on the special-relativistic resistive MHD (SRRMHD) evolution of the Orszag-Tang vortex over a range of resistivities spanning the Sweet-Parker and fast reconnection regimes. By embedding the governing equations as an additional loss term evaluated at finer temporal resolution than the available data supervision, the model learns dynamics at time steps where no simulation data is provided, enabling recovery of plasmoid formation that a data-only baseline trained on the same sparse snapshots fails to reproduce. To our knowledge, the present work is the first application of a physics informed neural operator to special relativistic resistive MHD, and the first to investigate the capability of such models to resolve plasmoid formation in SRRMHD.

In a second line of investigation, an OFormer-style Transformer Neural Operator is trained on the evolution of spine-sheath relativistic jets created with \texttt{BHAC}, in special-relativistic MHD (SRMHD).
The model is directly applied on the adaptive mesh, highlighting the need for linear attention due to long sequences.
The neural surrogate model is capable of capturing most of the major details, especially in early predictions.
To our knowledge, this constitutes the first application of a neural operator directly on a high resolution adaptive mesh refinement grid in the context of MHD simulations.

\end{abstract}

\section{Introduction}

General-relativistic magnetohydrodynamic (GR-MHD) simulations are indispensable for understanding high-energy astrophysical phenomena such as black hole accretion and relativistic jet launching~\cite{2019EHT_I,2019EHT_V}. Simulators like the Black Hole Accretion Code (\texttt{BHAC})~\cite{Porth2017, Olivares2019, Ripperda2019} can resolve these processes at high fidelity~\cite{2019BHAC_EHT}, but at substantial computational cost -- a single simulation may require thousands of CPU-hours, and systematic parameter studies demand hundreds of such runs. This expense limits the scope of explorable parameter spaces and hampers workflows that require rapid forward evaluations.

Neural operators offer a promising path toward alleviating this bottleneck. Unlike conventional neural networks, which learn mappings between fixed-dimensional vector spaces, neural operators approximate mappings between function spaces and are therefore invariant to the discretization of their inputs \cite{OperatorLearningMapsBetweenFunctionSpaces}. This property makes them natural candidates for learning surrogate models of PDE-governed systems. Recent work has demonstrated their effectiveness on a range of fluid-dynamical problems~\cite{li2021fourierneuraloperatorparametric,li2023physicsinformedneuraloperatorlearning,Rosofsky_2023,duarte2025spectrallearningmagnetizedplasma} -- yet their application to relativistic (resistive) MHD, with strong shocks, stiff source terms, and adaptive mesh refinement (AMR) grids -- remains largely unexplored.

This work investigates neural operator surrogates for two distinct \texttt{BHAC} simulation scenarios of astrophysical interest.

First, a physics informed Fourier Neural Operator (PINO) is trained on the special-relativistic \textit{resistive} MHD equations governing the Orszag-Tang vortex, a canonical benchmark for turbulent reconnection and plasmoid formation. This work investigates whether embedding the governing equations into the loss function can improve temporal generalization.

Second, an AMR-native neural operator based on the Galerkin Transformer architecture is developed to predict the propagation of relativistic jets directly on the irregular, multi-resolution grids produced by \texttt{BHAC}, avoiding the information loss inherent in regridding to uniform meshes. Together, these contributions represent a step toward practical, physics-consistent surrogate models for relativistic astrophysical simulations.


\section{Numerical Simulations}

The simulations of the Orszag-Tang vortex and the relativistic jets are both performed using the Black Hole Accretion Code (\texttt{BHAC}) \cite{Porth2017, Olivares2019, Ripperda2019}, a general-relativistic magnetohydrodynamic solver built on the MPI-AMRVAC framework \cite{keppens2012parallel, porth2014mpi}. \texttt{BHAC} is widely used to study black hole accretion, jet launching, and magnetic reconnection.
\texttt{BHAC} employs Adaptive Mesh Refinement (AMR): the mesh is refined in regions with fine structure in order to adequately represent them, while a coarse grid is used where sufficient. This saves on memory and computation. In practice, a moderate base grid can yield effective resolutions orders of magnitude higher.

Despite these efficiencies, the computational cost remains substantial. A single high-resolution run can take thousands of CPU-hours, and parameter studies require many such runs. This motivates the search for surrogate models that can approximate \texttt{BHAC}'s output at a fraction of the cost -- which is where neural operators come in.
Moreover, scalable models beyond the regime of the MHD approximation (e.g. to describe kinetic processes on sub-grid scales) may require new approaches.


\section{Neural Operator}

In contrast to neural networks, which approximate mappings between vector spaces $f : \mathbb{R}^n \to \mathbb{R}^m$, 
neural operators (NO) approximate operators $\mathcal{G} : \mathcal{U} \to \mathcal{V}$, mapping between function spaces \cite{OperatorLearningMapsBetweenFunctionSpaces}.
This removes the approximator's dependence on the input discretization, making the NO invariant to the applied resolution \cite{li2021fourierneuraloperatorparametric}.
It has shown great value in building surrogate models for a great variety of PDE systems, e.g. \cite{li2021fourierneuraloperatorparametric,li2023physicsinformedneuraloperatorlearning}.
Two different Neural Operators are investigated for different settings.
A Fourier Neural Operator for the Orszag-Tang Vortex and a OFormer-based operator for the relativistic jets.

\section{PINO for Resistive MHD}
\label{sec:pino}


This section presents a Physics Informed Fourier Neural Operator (PINO), that is trained to predict the evolution of the Orszag-Tang vortex in special-relativistic \textit{resistive} MHD (SRRMHD). The stiff source terms and fine current-sheet structures that characterize this regime make it both computationally expensive and a challenging test case for surrogate modeling. We investigate whether enforcing the governing equations through a PDE loss improves temporal generalization, up to the plasmoid-forming regime.

\subsection{Related Work}

Several recent works have explored neural operators as surrogate models for MHD simulations. Rosofsky \& Huerta~\cite{Rosofsky_2023} presented the first application of physics informed neural operators to 2D incompressible MHD, demonstrating accurate predictions for laminar flows with Reynolds numbers $\mathrm{Re}\leq250$. Bormanis et al.~\cite{Bormanis2024} investigated the Orszag-Tang vortex using physics-constrained convolutional neural networks with hard-coded divergence-free magnetic fields. Duarte et al.~\cite{duarte2025spectrallearningmagnetizedplasma} trained a purely data-driven FNO surrogate on Orszag-Tang vortices parameterized over viscosity and magnetic diffusivity. Other ML architectures for MHD have also been explored, including hybrid operator-diffusion frameworks for turbulent regimes~\cite{Kacmaz2025} and Flux Fourier Neural Operators for ideal MHD fluxes~\cite{kim2024neuraloperatorslearnlocal}.
Concurrent to this work, Cheung et al.~\cite{cheung2025reconstructingrelativisticmagnetohydrodynamicsphysicsinformed} applied physics informed neural networks (PINN) to ideal relativistic MHD; however, that work employs PINNs rather than neural operators and does not treat resistive effects.

 To our knowledge, the present work is the first application of a physics-informed neural operator to special relativistic resistive MHD, and the first to investigate the capability of such models to resolve plasmoid formation in SRRMHD.


\subsection{Numerical Simulation}
The Black Hole Accretion Code (\texttt{BHAC}) \cite{Porth2017, Olivares2019, Ripperda2019} is used to generate Orszag-Tang vortices at different resistivities.
Introducing resistivity into the SRMHD equations
results in a stiff source term in Ampère's law which make explicit solvers extremely inefficient.
Therefore, an IMEX scheme is used.
This means that the ideal MHD part is solved explicitly, while the stiff source term is solved implicitly. \cite{Ripperda2019}
Additionally, a entropy-based recovery scheme is employed in \texttt{BHAC} to make inversion more stable.

The special-relativistic resistive MHD equations are provided in appendix \ref{sec:equations}.

\subsection{Oszag-Tang vortex}
The Orszag-Tang vortex \cite{Orszag_Tang_1979} is a toy model for turbulent SRRMHD flow.
It is well suited for investigating strong shocks and plasmoid formation.
The specific setup for the Orszag-Tang vortex is more detailed in appendix~\ref{sec:app_OTV}.
A $2.5$ dimensional Cartesian grid is used, with a periodic spatial domain $(x,\,y) \in [0,\,2\pi]^2$, and $t\in[0,\,10]$, using natural units.

In order to get physically meaningful results, the maximum refinement level needs to be chosen large enough for the simulation to converge. For this, Orszag-Tang vortices at different AMR levels are compared in $\langle B^2\rangle=\iint_VB^2 \, dx\,dy / \iint_Vdx\,dy$.
Convergence is claimed when $\langle B^2\rangle$(t) does no longer change with successive increase of the AMR level, according to \cite{Ripperda_2020}.

The investigated resistivity range is $\eta\in [10^{-4},\,10^{-3}]$. For this an AMR level of $5$ was found sufficient.
Plasmoid development can be observed for $\eta\to 10^{-4}$.
Those plasmoids form through violent tearing instabilities of the current sheets. They are of particular interest to the Astrophysical community because of their conjectured role in powering frequent X-ray and near infrared flares from Sgr A*, our Galaxy's black hole. \cite{Ripperda_2020}
Towards $\eta=10^{-3}$ the current sheets do not get thin enough for plasmoid formation.
The investigated resistivity range thus covers both the Sweet–Parker regime as well as the fast reconnection regime.

The investigated and predicted quantities for the Orszag-Tang vortex are $(u_x,\,u_y,\,B_x,\,B_y,\,\rho,\,E_z,\,p)$, where subscripts $x$ and $y$ denote the in-plane Cartesian components, $u$ being the velocity, $B$ the magnetic field, $\rho$ the rest mass density, $E_z$ the $z$-component of the electric field, and $p$ the pressure.
These quantities fully quantify the state of the resistive SRMHD system through the equations in the appendix~\ref{sec:equations}.


\subsection{FNO}

A Fourier Neural Operator (FNO) \cite{li2021fourierneuraloperatorparametric} was chosen for this task because of its unique properties and its success on comparable tasks, e.g. \cite{li2021fourierneuraloperatorparametric,li2023physicsinformedneuraloperatorlearning,Rosofsky_2023,duarte2025spectrallearningmagnetizedplasma}.
As a neural operator it is resolution invariant, in both spatial and temporal dimension.
Its specific architecture is explained further in the appendix~\ref{sec:app_FNO}.
The simulation data from \texttt{BHAC} is converted to a uniform grid for the Orszag-Tang vortex.

The FNO maps the initial conditions and the resistivity parameter $\eta$ to the full temporal evolution of $(u_x,\,u_y,\,B_x,\,B_y,\,\rho,\,E_z,\,p)$ over the $2+1\,$D domain. 
Crucially, the temporal dimension is treated on equal footing with the spatial dimensions -- the Fourier layers operate over the full $2+1\,$D spatiotemporal domain. This allows the network to learn the complete dynamical evolution in a single forward pass rather than autoregressively stepping through time.
This has proven to improve the model quality especially at late time steps, but comes at the cost of higher GPU memory requirement.

\begin{figure}[t]
    \centering
    \includegraphics[width=\textwidth]{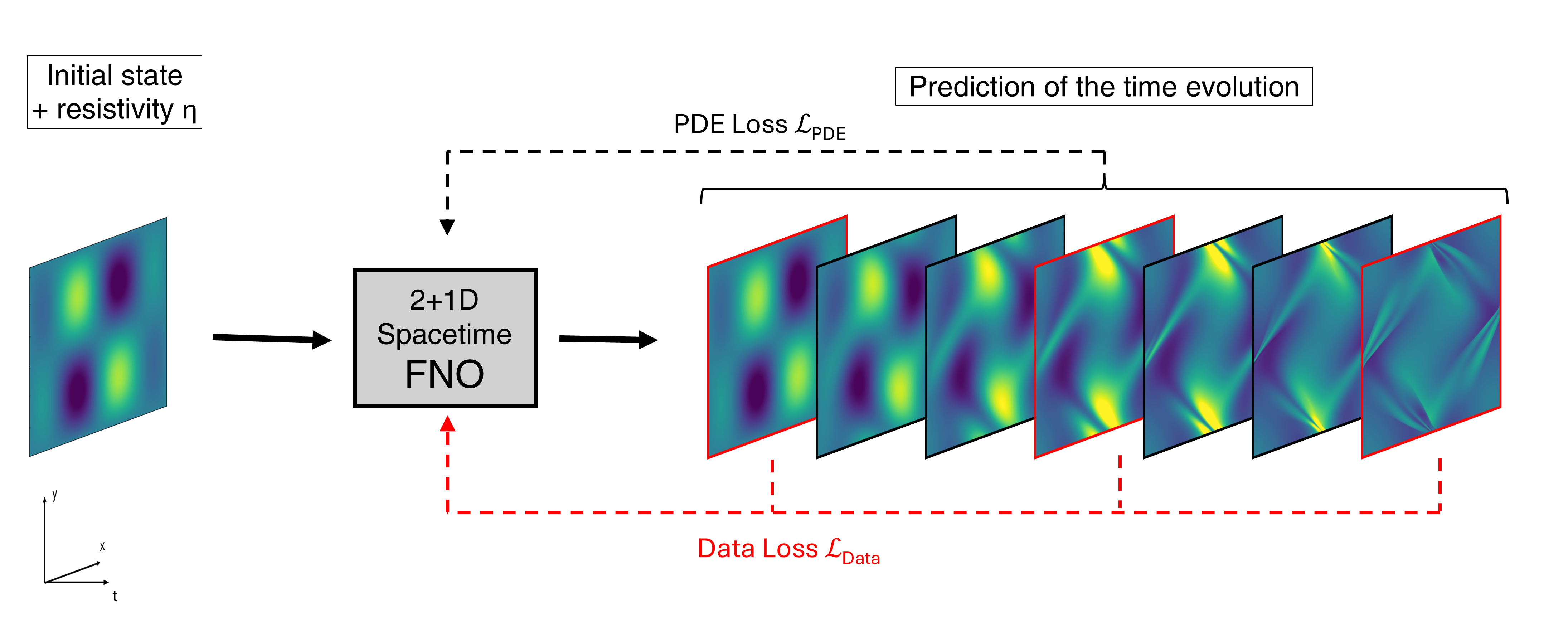}
    \caption{Illustration of the used setup. The Fourier Neural Operator (FNO) received an initial state and a resistivity $\eta$ as input. It predicts the Oszag-Tang evolution. The data Loss $\mathcal{L}_{data}$ is only applied at a few timesteps (red frames), whereas the PDE Loss $\mathcal{L}_{PDE}$ is enforced on a much finer temporal resolution (red and black frames). This proves to be effective in helping the model learn at timesteps without data supervision (black frames). }
    \label{fig:PDEvsData}
\end{figure}

The FNO setup is illustrated in \autoref{fig:PDEvsData}, which is later also used to explain the physics informed loss.

\subsection{Physics Informed}

Training good surrogate models typically requires large volumes of high-resolution simulation data, which are themselves expensive to generate. Physics informed methods address this limitation by embedding governing equations directly into the learning objective, reducing dependence on labeled data. This work investigates the efficacy of incorporating the governing SRRMHD equations (appendix~\ref{sec:equations}) into the FNO loss function.


Because the FNO predictions $(u_x,\,u_y,\,B_x,\,B_y,\,\rho,\,E_z,\,p)$ fully determine the state of the SRRMHD system, the governing equations (appendix~\ref{sec:equations}) can be evaluated directly on the model output. For this spectral derivatives are used. The PDE residuals are then introduced as an additional physics informed loss term $\mathcal{L}_{PDE}$, supplementing the data-driven loss $\mathcal{L}_{data}$.
The implementation is based on the PhysicsNeMo framework \cite{physicsnemo}, which was adapted for the present setting.

To test the efficacy of this physics loss, the model is trained on a data set with poor temporal resolution, while its performance is tracked on the same data set at $8\times$ higher temporal resolution. The PDE loss is applied at this same $8\times$ higher temporal resolution. This setup is illustrated in \autoref{fig:PDEvsData}, where the PDE loss acts at $3\times$ higher temporal resolution for clarity

\subsection{Results}
In total 29 simulations are used, where the resistivity $\eta$ is chosen randomly\footnote{The random selection was checked to ensure reasonably uniform coverage of the full range, including the edges.} in the range  $\eta \in [10^{-4},\,10^{-3}]$. The data is split into training and validation samples; 23 and 6 samples respectively.
The full physical time range of $t \in [0,\,10]$ is represented in the simulation by $401$ evenly spaced slices.

All results presented of the model's performance are measured on validation samples; meaning interpolated, the dynamics at unseen resistivities.

Training is done on three NVIDIA L40 in a Distributed Data Parallel (DDP) configuration.

\subsubsection{Data Supervision}

The model's performance is investigated for pure data supervision.
For GPU memory reasons the Orszag-Tang evolution is split into several time domains, each treated independently.

\begin{figure}[H]
    \centering
    \includegraphics[width=\textwidth]{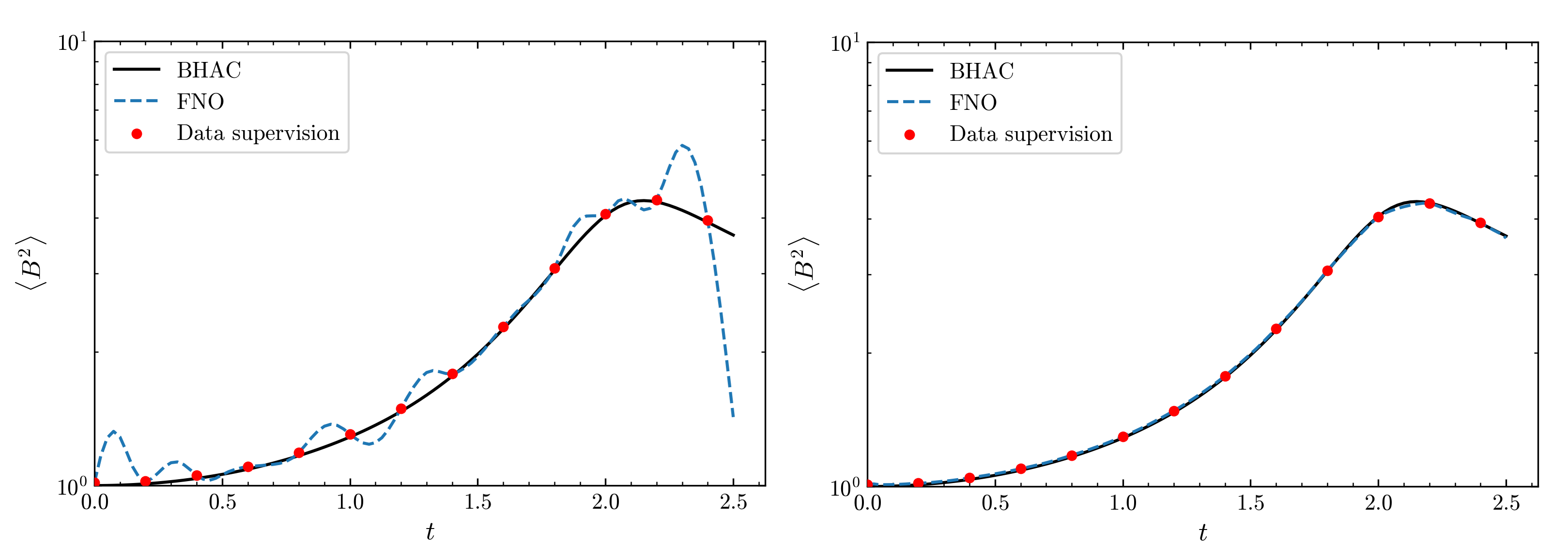}
    \caption{Domain averaged magnetic energy density $\langle B^2 \rangle$ as a function of time at a resistivity of $\eta=5.59\cdot10^{-4}$. Training without (left) vs. with (right) PDE constraint. The are enforced on an 8 times finer time-grid than the data loss. A clear improvement in the model's performance can be seen.}
    \label{fig:B2_vs}
\end{figure}

\paragraph{Time $t \in [0,\, 2.5]$:}
No plasmoids are created in this time domain. Yet strong shocks are exhibited.
A spatial resolution of $256^2$ is used. The data loss is applied only to every 8th simulated time step, while performance is evaluated at \textit{every} time step, so that the model's ability to interpolate to intermediate times can be assessed. The FNO learns the training snapshots well and generalizes accurately to unseen resistivities, yet it performs poorly at the intermediate time steps on which it received no data supervision -- it does not generalize to higher temporal resolution. This is visible on the left in \autoref{fig:B2_vs}. The fidelity of this is dependent on the resistivity. $\eta \to 10^{-4}$ shows slightly larger deviation at $t\to2.5$ than~\autoref{fig:B2_vs}, at $\eta = 5.59\cdot10^{-4}$. This is likely due to numerical resistivity, because the PDE residuals are evaluated on a relatively coarse grid, introducing discretisation errors.

\paragraph{Time $t\in [9.5,\,10]$:} This is plasmoid regime. The spatial resolution is increased to $512^2$ to better resolve the fine structure of the current sheets, and the current density $J_z$ is computed from the model outputs as a diagnostic. As in the early-time regime, the FNO is trained on every 8th time step and evaluated at all intermediate steps. The results are qualitatively similar: the model reproduces the large-scale structure adequately but fails to capture fine-scale features -- in particular, plasmoids are not reliably predicted at unseen time steps. This is illustrated in the central panels of \autoref{fig:Jz_01}, and in the appendix \autoref{fig:Jz_00} and \autoref{fig:Jz_03}.

\subsubsection{Physics Informed Training}

The most stable method for introducing the PDE loss was found to be as follows. The model is trained with purely data supervision for 100 epochs. Then the PDE loss weight is increased linearly with the epoch count, with the growth rate stepping up at epochs $[500,\, 700,\,1100]$.

\paragraph{Time $t \in [0,\, 2.5]$:} The PDE loss is applied at a finer temporal resolution. The data supervision from before is kept, but additionally the PDE loss applied on intervening time steps: each time steps has the PDE loss applied while the data loss is only applied at every $8^{\text{th}}$ step.
The result from this can be seen in the right panel of \autoref{fig:B2_vs}. A clear improvement is visible at time steps without data supervision.
Further comparisons of the model's predictions without and with PDE constraints are shown in \autoref{fig:E3_vs}, and in the appendix \autoref{fig:B2_slice_vs} and \autoref{fig:By}.

\begin{figure}[H]
    \centering
    \includegraphics[width=0.75\textwidth]{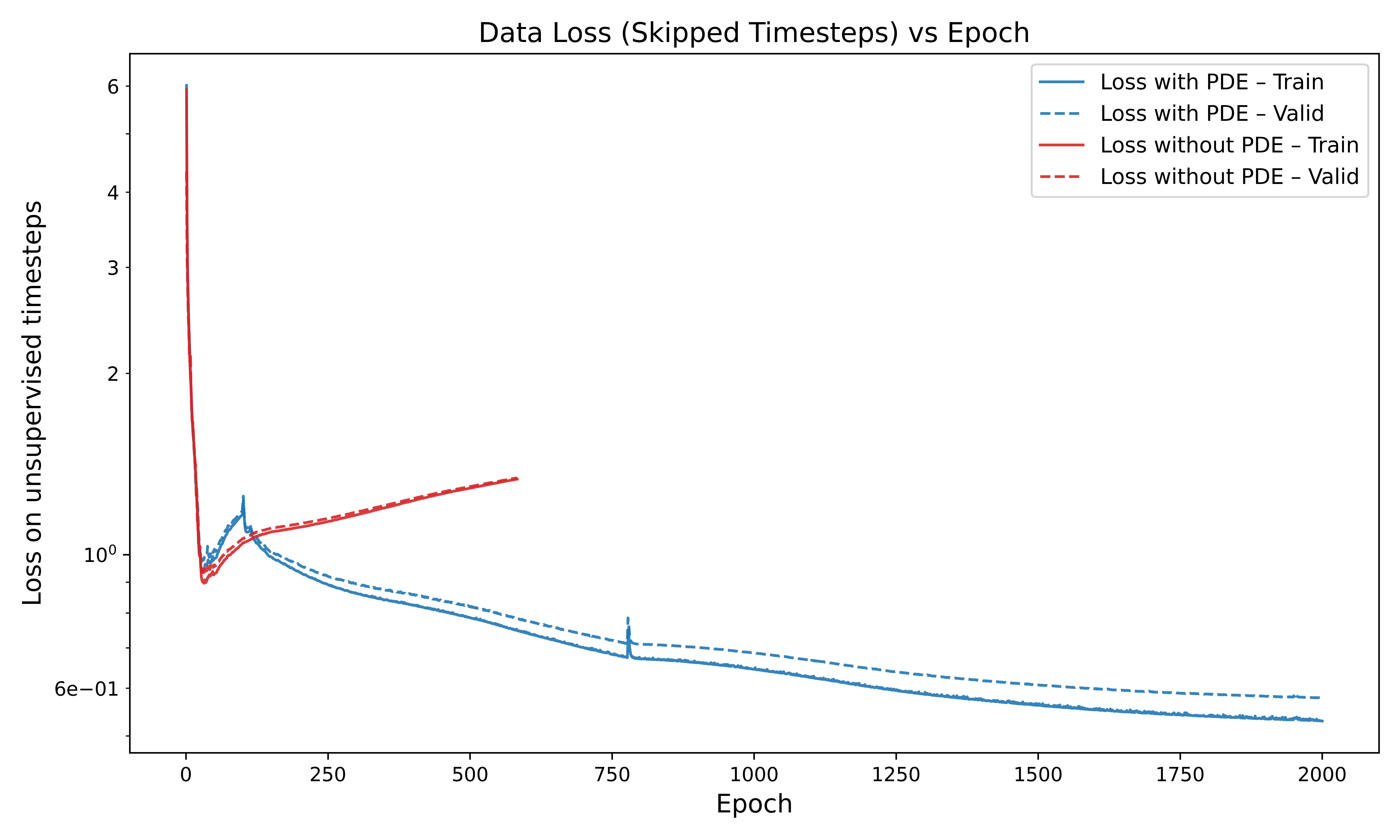}
    \caption{The relative L$_2$-loss (evaluated on \texttt{BHAC} data) on the time steps without data supervision, without (red) and with (blue) PDE constraint. Without PDE constraints the model fails to generalize well to unseen time steps. For the blue curves the PDE loss is activated at epoch $100$.}
    \label{fig:Loss_vs}
\end{figure}

The corresponding loss curves (\autoref{fig:Loss_vs}) confirm that: without PDE constraints the loss at time steps without data supervision remains high and even trends upward, whereas it decreases steadily once the PDE loss is included.

\begin{figure}[H]
    \centering
    \includegraphics[width=\textwidth]{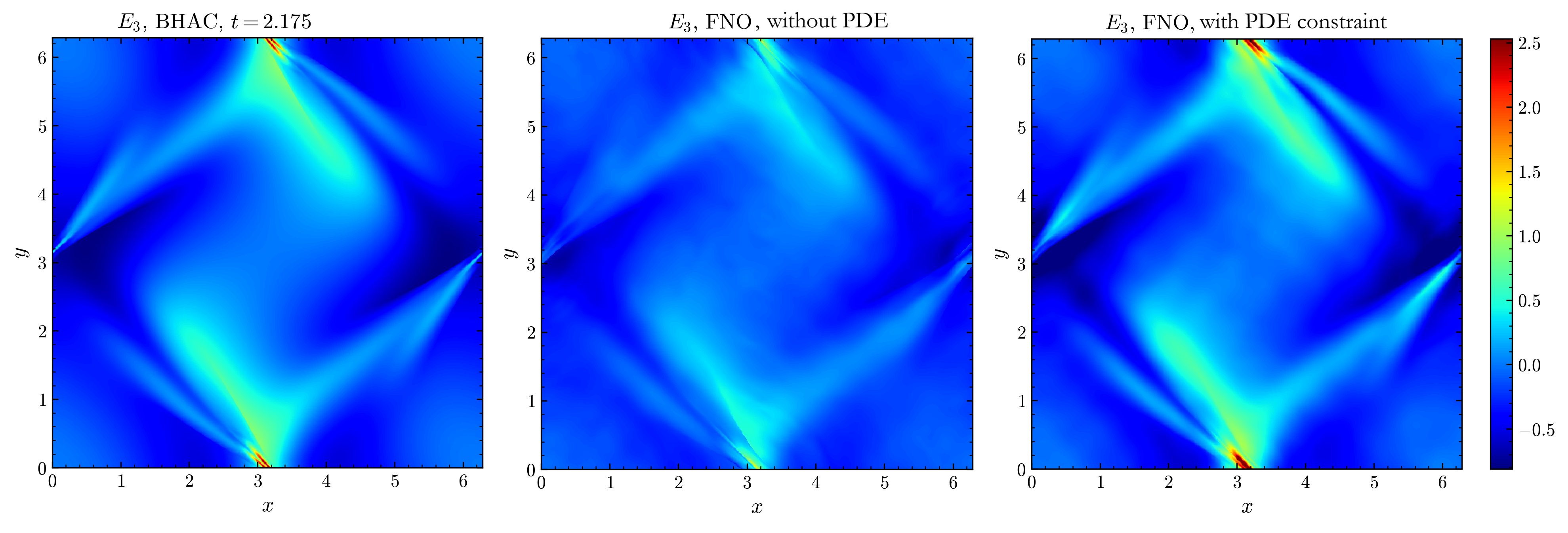}
    \caption{Electric field $E_z$ at a timestep without data supervision. Ground truth (left), the model's interpolation without PDE constraints (middle) and the model's predictions with PDE constraints enforced (right). Including the PDE loss helps the model make better predictions}
    \label{fig:E3_vs}
\end{figure}

\paragraph{Time $t\in [9.5,\,10]$:}
Again the PDE loss is enforced on a $8\times$ finer time mesh than the data. A clear improvement is visible in \autoref{fig:Jz_01}, and in the appendix \autoref{fig:Jz_00} and \autoref{fig:Jz_03}. Plasmoids are formed in the physics informed model predictions, which were not present in the model predictions without physics information.

\begin{figure}[H]
    \centering
    \includegraphics[width=\textwidth]{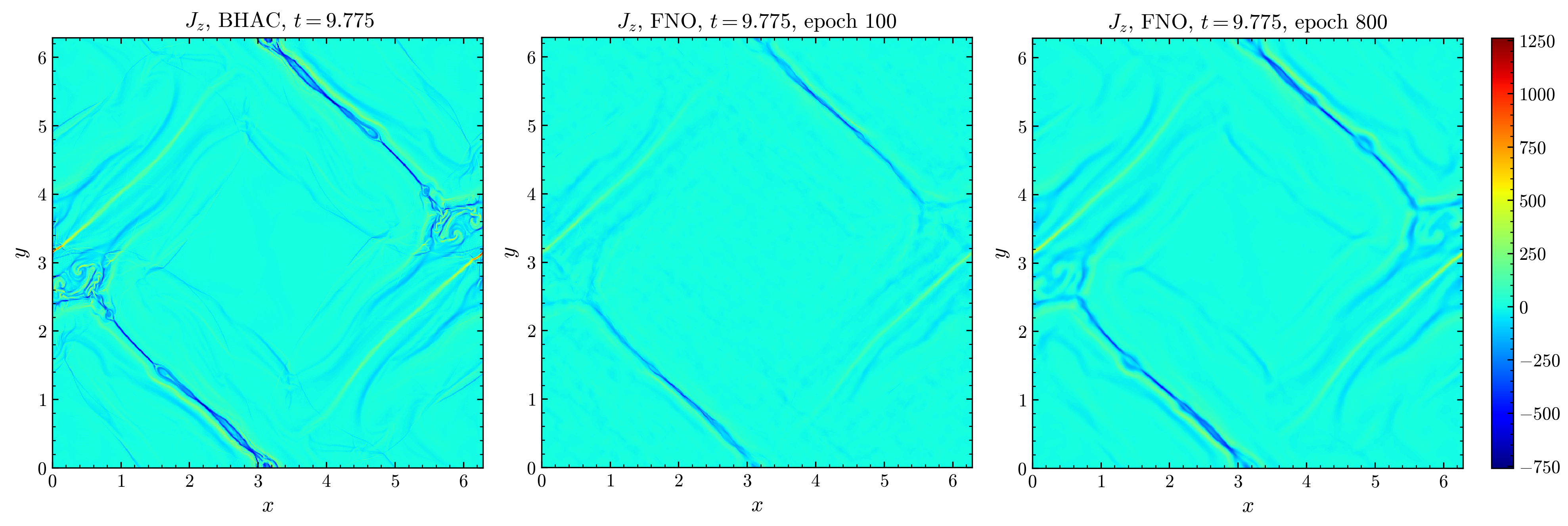}
    \caption{Electric current density $J_z$ at a representative timestep without data supervision. $\eta = 1.08\cdot10^{-4}$. Ground truth (left), the model's interpolation without PDE constraints (middle) and the model's predictions with PDE constraints enforced (right). Plasmoids are visible in the physics informed model, that were not present in the model without physics information.}
    \label{fig:Jz_01}
\end{figure}

\subsection{Discussion}

This work demonstrates that incorporating the governing PDEs into the loss function significantly improves the model's accuracy at time steps lacking data supervision. Most notably, the physics informed model successfully resolves plasmoid structures that the purely data-driven model fails to capture. To our knowledge, this constitutes the first demonstration of a neural operator reproducing plasmoid dynamics in relativistic resistive MHD.\\
The method also shows some limitations.

i) Increasingly high training resolutions for finer structures are needed for high fidelity of the PDE loss. This limits the $2+1\,$D spacetime FNO approach in GPU memory.

ii) The stiffness in Ampère's law induces dynamics on very short timescales, imposing strict temporal resolution requirements analogous to those in classical solvers: the PDEs must be evaluated at a sufficiently fine time scale for the resulting gradients to remain meaningful.
This effect becomes more pronounced at smaller resistivities, constraining the present method in the low-resistivity regime.

iii) When choosing a lower training resolution than the highest AMR level in the simulation (the effective resolution), artifacts from downsampling are introduced.
These artifacts affect both the standard field representation and the PDE residuals through the derivatives.

iv) \texttt{BHAC} is built to satisfy the governing PDEs on conserved variables, whereas this work operates on primitive variables. Evaluating the PDEs on the primitive representation yields high residuals at shock fronts -- discontinuities in the solution -- meaning the training data itself does not satisfy the imposed PDEs well.

Together, limitations iii) and iv) complicate the monitoring of model performance: an increase or decrease in the metrics evaluated on the simulated data may equally indicate an improvement or a degradation of the model's physical fidelity.

Regarding limitation iv), reformulating the physics loss in weak or integral form -- as explored for conservation laws in wPINNs~\cite{deryck2022wpinnsweakphysicsinformed,shocksPINN} and IPINNs~\cite{rajvanshi2024integral} -- offers a promising path forward, since these formulations naturally accommodate discontinuities such as shocks.

\section{AMR-native Neural Operator on GR-MHD}

Relativistic jets simulated with \texttt{BHAC} rely on adaptive mesh refinement to efficiently predict their evolution in time and space.
This section identifies a neural operator surrogate alternative that can successfully predict the special relativistic magnetohydrodynamic physics.
As AMR leads to drastically changing scales and long sequences of sampled tokens, the challenge is to find a NO that bypasses these constraints and still provide competitive performance and runtime advantages compared to the simulation counterparts.

\subsection{Related Work}

Modern Computational Fluid Dynamics (CFD) simulations, such as \texttt{BHAC}, \texttt{KHA\-RMA}~\cite{prather2025kharma}, or \texttt{OpenFOAM}~\cite{jasak2007openfoam, jasak2009dynamic}, rely on Adaptive Mesh Refinement (AMR) to achieve high precision with computational efficiency.
Despite being theoretically resolution invariant, many modern NO approaches cannot be applied directly to these non-uniform AMR meshes.
For instance the FNO~\cite{li2021fourierneuraloperatorparametric} requires a regular Cartesian grid (cf. \autoref{sec:pino}).
Variations like the Spherical FNO (SFNO)~\cite{bonev2023spherical} extend this to spherical geometries via generalized Fourier transforms~\cite{driscoll1994computing}, while Geo-FNO~\cite{li2023fourier} attempts to learn grid deformations from irregular grids to a regular latent space where FNOs can operate.

Graph Neural Operators (GNO)~\cite{li2020neural} rooted in Graph Neural Networks~\cite{scarselli2008graph, kipf2016semi} construct a radius graph in the ball neighborhood of Euclidean space, making them natively compatible with irregular meshes.
The Geometry-informed Neural Operator (GINO)~\cite{li2023geometry} combines a GNO encoder/decoder to bridge irregular input meshes with FNOs as processors in a regular latent space.
While FNO processors are generally preferred for their efficiency over similarly scaled GNOs, our pre-study revealed that GINO yields heavily over-smoothed predictions when applied to AMR data. 
We attribute this to the combination of fixed-radius graph construction necessary for resolution-invariance and the inherent resolution bottleneck of the regular latent grid, which fails to preserve the high-frequency features captured by the refined AMR regions.
The application of Nyström sampling~\cite{nystrom1930praktische, li2020neural} does not alleviate this issue, as the required uniform sampling must be performed prior to ball construction, effectively discarding the AMR's local density benefits. 
Furthermore, we argue that the common implementation of a \textit{max-k} parameter in radius graph construction-either used for efficiency or imposed by CUDA kernel optimization-breaks strict resolution invariance and contributes to the poor scaling on large meshes observed by \citeauthor{zhou2026transolver}~\cite{zhou2026transolver}.
The Codomain Attention Neural Operator (CoDA-NO)~\cite{rahman2024pretraining} faces similar limitations; while it innovates by tokenizing physical variables rather than spatial locations, it still relies on a GINO-like mapping for irregular geometries, inheriting the same spatial bottleneck.

Recently, transformer-based approaches have emerged as a robust alternative.
\citeauthor*{calvello2025continuum}~\cite{calvello2025continuum} demonstrated that attention~\cite{vaswani2017attention} and ViT~\cite{dosovitskiy2020image} can be extended to the operator framework.
To handle irregular domain discretizations without intermediate regular grids, the Universal Physics Transformer (UPT)~\cite{alkin2024universal} encodes arbitrary point clouds into a unified latent space where attention is applied, enabling continuous space-time querying via a decoder.
Similarly, iterations such as the Mesh-Informed Neural Operator (MINO)~\cite{shi2025mesh} and the Geometry-Aware Operator Transformer (GAOT)~\cite{wen2025geometry}, utilize GNO-based tokenizers with cross-attention or ViT processors and with the latter using attention weights to refine the GNO integral transform.
Other architectures, such as the Transolver family~\cite{wu2024transolver, luo2025transolver++, zhou2026transolver}, attempt to bypass GNO scaling difficulties through token "slicing" to group mesh points into physics-aware tokens, reducing the attention complexity from $O(N^2)$ to linear.
Explicitly targeting AMR data, the recent AMR-Transformer~\cite{xu2025amr} directly utilizes the hierarchical multi-way tree structure of AMR meshes as a tokenizer.
By coupling this with constraint-aware pruning based on fluid properties, it selectively drops tokens in calm regions to make the standard attention mechanism computationally feasible.

However, heuristic token pruning and discrete tree formulations often restrict models from acting as true resolution-invariant continuous operators.
To maintain AMR-native resolution without the smoothing effects of regular-grid proxies while scaling to massive token sequences, linear attention mechanisms show the greatest potential.
By utilizing a Galerkin-style linear attention~\cite{cao2021choose}, OFormer~\cite{li2022transformer} avoids the quadratic $\text{Softmax}(\cdot)$ memory bottleneck while remaining compatible with disparate input and query grids. 
Based on these considerations and the failures of hybrid models observed in our preliminary work, we present OFormer as a viable AMR-native model and demonstrate its effectiveness on a custom relativistic jet dataset governed by SRMHD.

%

\subsection{Methodology}

The model we use for this is based on the OFormer~\cite{li2022transformer} architecture, which in turn is based on an attention integral based on the Galerkin-style transformer \cite{cao2021choose}.
By removing $\text{Softmax}(\cdot)$ from standard attention \cite{vaswani2017attention} and changing the order of matrix multiplication, memory consumption drops to $O(nd^2)$, with $n$ as sequence length and $d$ as the model embedding dimension:
\begin{equation}
    z = \text{Attn}_{\text{g}}(y) := \frac{1}{n}Q\left(\tilde{K}^T\tilde{V}\right)
\end{equation}
where $z$ is the output of Galerkin-style attention $\text{Attn}_{\text{g}}(y)$ of input $y$, $n$ is the sequence length, and $Q = y\cdot W_Q$, $\tilde{K} = \text{Ln}(y\cdot W_K)$, and $\tilde{V} = \text{Ln}(y \cdot W_V)$ are the attention matrices with $\text{Ln}(\cdot)$ as layer normalization~\cite{ba2016layer} to mimic the regularizing effect of $\text{Softmax}(\cdot )$.

This standard implementation uses the same input and output grid and is only applicable to regular grids as integration weights are not considered.
To this end, \citeauthor*{li2022transformer}~\cite{li2022transformer} propose the use of an Encoder-Decoder-Style Transformer NO with a generalized, weighted Galerkin-style attention integral.
The resulting model is similar to the weighted attention integral and Encoder-Decoder NOs presented by \citeauthor{berner2025principled} \cite{berner2025principled}.
We formulate self-attention $\text{Attn}_{g}'(y)$ as follows, which is necessary to support long sequences like those regularly found with AMR grids:
\begin{equation}
    z' = \text{Attn}_\text{g}'(y) := Q\left(\tilde{K}^T\left(\tilde{V}\odot w\right)\right)
\end{equation}
with $w_i = \frac{v_i}{\sum_{j=1}^n v_j}$ as the normalized quadrature weight of the AMR block $i$ with block-volume $v_i$.
To map to a different target grid, we also introduce weighted cross-attention $\text{Attn}_{g,\times}(y_\times, y)$ for the Decoder in a similar way:
\begin{equation}
    z_\times' = \text{Attn}_{\text{g},\times}'(y_\times, y) := Q\left(\tilde{K}^T_\times \left(\tilde{V}_\times \odot w_\times\right)\right)
\end{equation}
with $y_\times$ and $w_\times$ as the output of the Encoder network and the integration weights on the original grid, and the cross-attention matrices $\tilde{K}_\times = \text{Ln}(y_\times \cdot W_{K,\times})$ and $\tilde{V}_\times = \text{Ln}(y_\times \cdot W_{V, \times})$.
For better efficiency, we use root mean squared layer normalization (RMS layer norm)~\cite{zhang2019root} instead of standard layer normalization or instance normalization.

Apart from the attention integral transforms, the rest of the Encoder and Decoder network are similar to vanilla transformers without causal masking.
Both networks keep the skip connections and the classic two-layer feed-forward multi-layer-perceptron (MLP) with $\text{GeLU}(\cdot)$ \cite{hendrycks2016gaussian} as activation function.
For positional encoding, we use Rotary Embeddings~\cite{su2024roformer}.
The complete Encoder and Decoder layers are also represented in the Appendix in \ref{sec:app_oformer-architecture}.

For the final model, the input to the Encoder is $[f_x, p_x, g]$, the concatenated vector, consisting of input function $f_x$ evaluated at coordinates $p_x$ and potential global features $g$, is lifted with an MLP before the Encoder layers.
The Decoder layers are ingested with a lifted version of $[p_y, g]$ where $p_y$ are the query positions of the output function, and additionally the latent output of the Encoder $\bar{f}_{y,\times}$ in the cross-attention.
Finally, the latent output of the Decoder is projected to the target $f_y$ using a final MLP.




\subsection{Experiments}

\subsubsection{Relativistic Jet Data set}
For testing the effectiveness of different models on AMR grids, we created a custom data set of spine-sheath relativistic jets with \texttt{BHAC}.
The data is simulated onto a 2D grid with a base resolution of $256^2$ and $4$ additional refinement levels, resulting in an effective resolution of $4096^2$.
This results in a feature vector with six primitive variables, as the remaining \texttt{BHAC} variables can either be computed from these primitives or are $0$ in the 2D case; a summary of all available variables can be found in the Appendix in \ref{sec:app_jet-vars}.
Additionally, we use the simulation time $t$ and three changing input parameters of the jet as global parameters.
The final data set contains $448$ simulations with $T=50$ consecutive sample pairs $(s_{t-1}, s_{t}) \forall t\in [1, T]$ each, where $s_t$ is the sample at time $t$.
Of these $448$ simulations, we only use $421$ for training and completely isolate $27$ random simulations for final evaluation.
Additional information on the creation of the data set can be found in the Appendix in \ref{sec:app_jets}.

\subsubsection{Training}
Our model is initialized with an embedding dimension of $128$ and $12$ Encoder and Decoder layers where each attention uses $8$ attention heads with a head dimension of $16$.
We use a small dropout of $0.05$ to promote generalization and apply additional Nyström approximation~\cite{nystrom1930praktische, li2020neural} to uniformly sample $150\text{k}$ points during training.
The latter can also be seen as a kind of position augmentation as the model quickly needs to learn how to accurately encode positions efficiently.
This constant is also chosen because it ensures that the model can be trained on a GPU with $48\text{GB}$ VRAM.

We train our model with an $80/20$ training/validation split and an effective mini-batch size of $48$ for $256$ epochs using the relative $\text{L}_2$-loss~\cite{kovachki2023neural} and the AdamW optimizer~\cite{loshchilov2017decoupled} with a weight decay of $10^{-2}$.
Due to the different magnitudes of the channels, we note that we take the relative $\text{L}_2$-loss per channel first, before applying mean aggregation.
Our learning rate is controlled by a cosine learning rate scheduler with linear warmup for $8$ epochs, an initial learning rate of $10^{-7}$, a peak learning rate of $10^{-4}$, and a terminal learning rate of $10^{-6}$.

Due to the difference in magnitude of the input fields, we apply $\log(\cdot)$ transformation to $p$ and $\rho$ and normalize all data with \textit{Z-Standard} normalization with statistics calculated on the whole data set, except for the positions that we normalize to a range of $[0, 1]$ as this makes it easier to rescale them for Rotary Embedding.

We train the model with next-timestep-prediction, i.e., the output at time $t$ and query positions $p_t$ is produced from the inputs at time $t-1$ sampled at positions $p_{t-1}$.

\subsubsection{Results}


We continue with the model achieving the lowest $\text{L}_2$-loss on the evaluation split, which is reached after $204$ epochs.
The results are produced on the excluded $27$ simulations to test the true generalization capabilities of the model and without Nyström sampling, i.e., taking into account the entire AMR mesh.

From the results that are captured in \autoref{tab:transformer-results} we can see that the model is able to successfully learn the overall dynamics of the simulation.
The identity baseline that is given as a comparison is notably surpassed, and the model even gets relatively close to the performance of an overfitted model, which serves as an empirical lower bound for the achievable error given the current architecture and problem constraints.
More information on the baselines can be found in the Appendix in \ref{sec:app_oformer-baselines}.

\begin{table}[!ht]
    \centering
    \caption{Error of OFormer against Identity baseline and overfitted OFormer for comparison.}
    \begin{tabular}{|l|c|c|}
        \hline
        Model & $\text{L}_2$ error & $\text{L}_\infty$ error \\\hline
        OFormer & $0.245$ & $1.028$ \\\hline
        Identity & $0.372$ & $1.631$ \\\hline
        \textit{Overfitted OFormer} & $\mathit{0.092}$ & $\mathit{0.412}$ \\\hline
    \end{tabular}
    \label{tab:transformer-results}
\end{table}

\begin{figure}[H]
  \centering
  \includegraphics[width=0.8\linewidth]{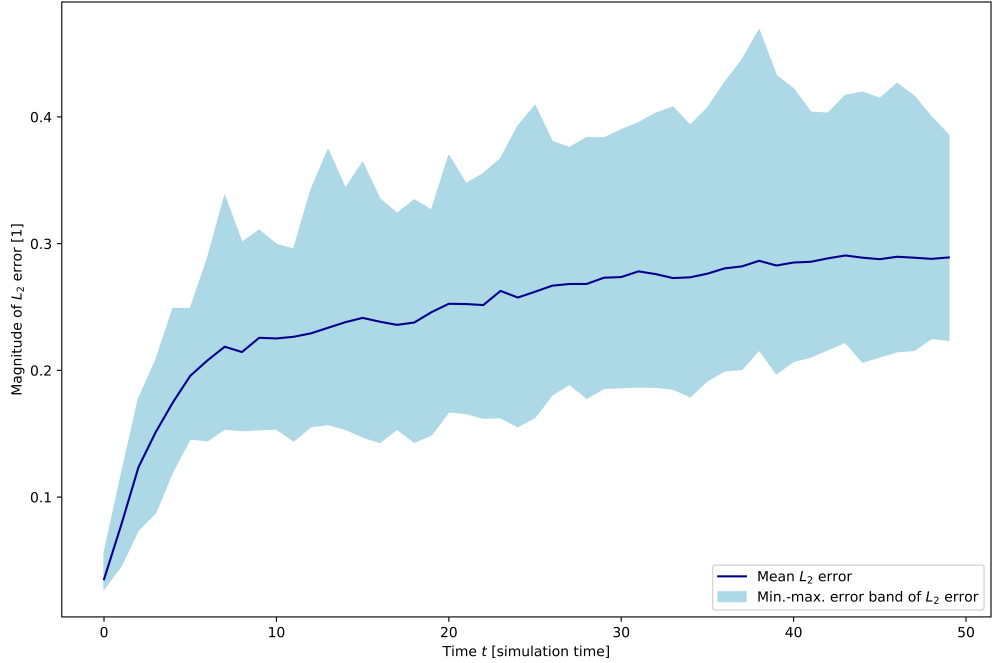}
  \caption{Evolving $\text{L}_2$ error of the OFormer on the test set. After low initial errors, the model starts to produce worse predictions for later timesteps when finer details are missed. These metrics are calculated on the $27$ test simulations.}
  \label{fig:transformer-over-time}
\end{figure}

However, we also observe that the performance is not uniform across simulation time.
As shown in \autoref{fig:transformer-over-time}, the prediction error starts low and rises for later time steps.
The quality gap between the best and worst performing example for each time step is also increasing, which hints that some simulations can be recreated better than others.

\begin{figure}[H]
    \centering
    \includegraphics[width=\linewidth]{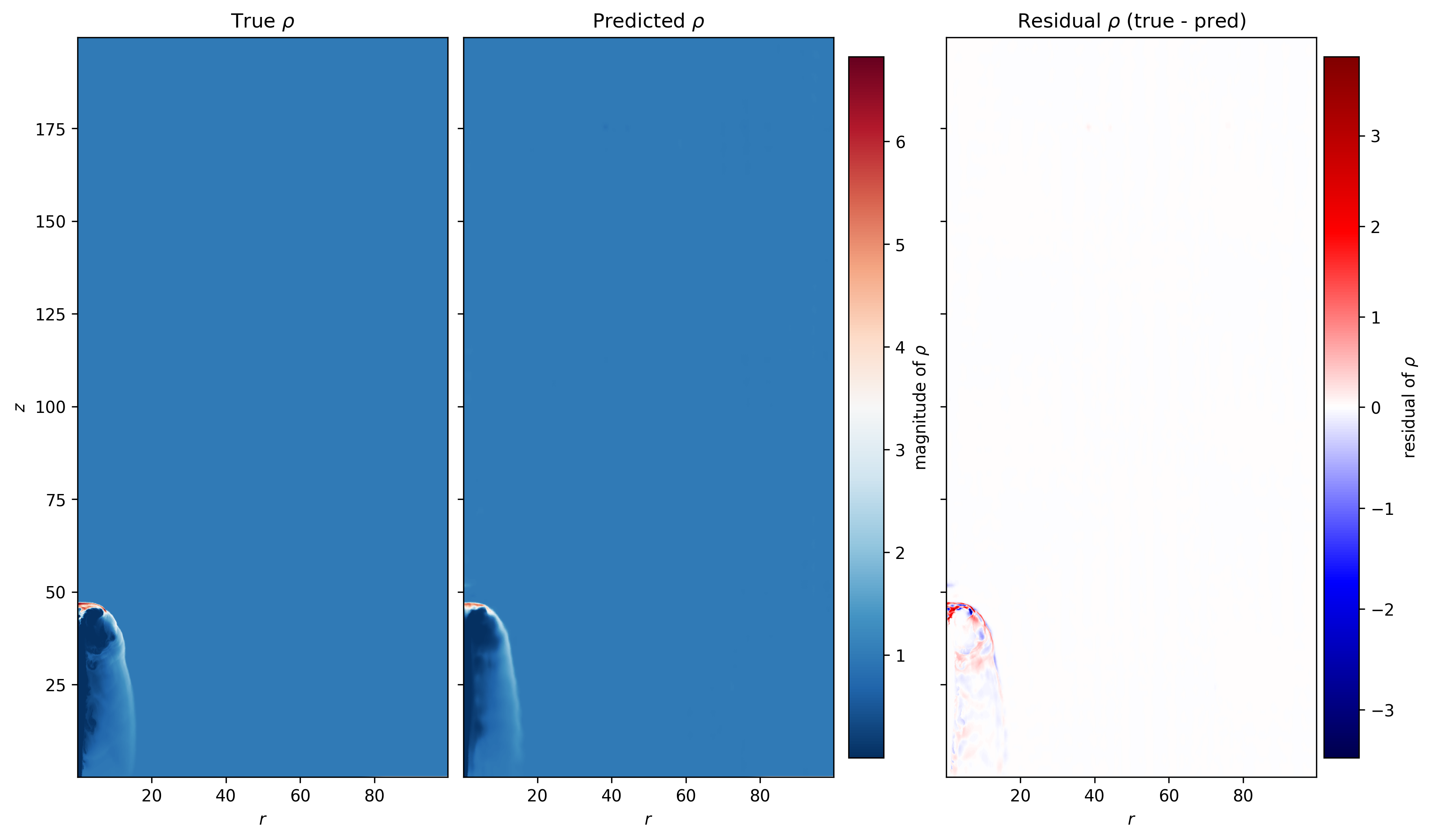}
    \caption{Prediction of $\rho$ at an early timestep ($t=5$). The parameters for the jet are $d_k=4$, $v_b=0.995$, $\eta=0.006$. All major details are successfully reconstructed including the top shockwave, while finer details like swirls cannot be recreated perfectly. ($\text{L}_2=0.212$, $\text{L}_\infty=0.697$, $\text{L}_{2, \rho}=0.185$)}
    \label{fig:transformer-early-predictions}
\end{figure}

Looking at an early prediction of $\rho$ shown in \autoref{fig:transformer-early-predictions}, we can see that the jet did not fully evolve, yet.
This makes it easier for the model to accurately predict the physics of the simulation allowing it to even capture shock waves.

\begin{figure}[H]
    \centering
    \includegraphics[width=\linewidth]{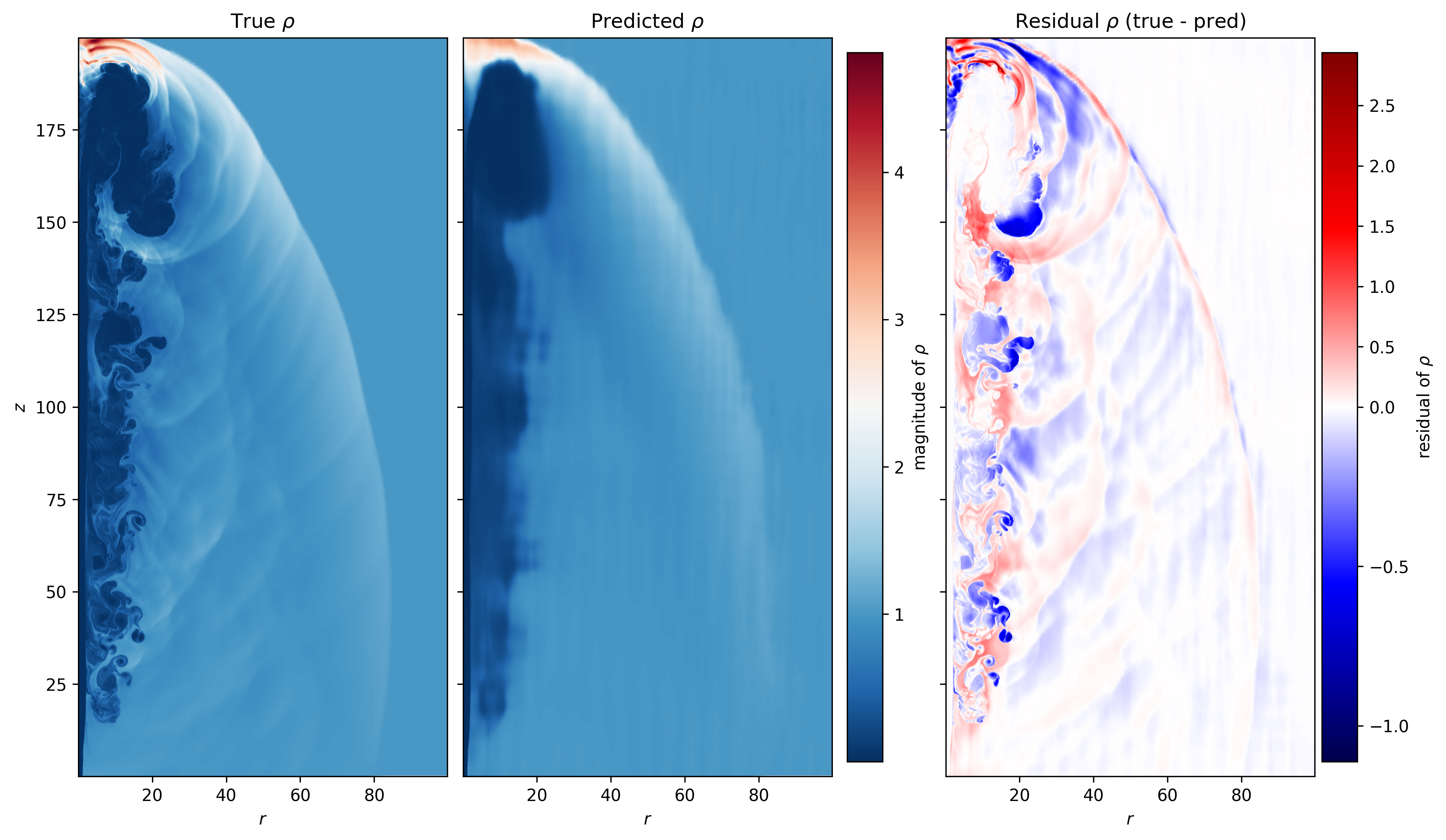}
    \caption{Prediction of $\rho$ at a later timestep ($t=40$). The parameters for the jet are $d_k=4$, $v_b=0.995$, $\eta=0.006$. Major details are mostly reconstructed while some finer details are missing or oversmoothed. ($\text{L}_2=0.366$, $\text{L}_\infty=0.729$, $\text{L}_{2, \rho}=0.396$)}
    \label{fig:transformer-late-prediction}
\end{figure}

In contrast, later predictions of $\rho$ as shown in \autoref{fig:transformer-late-prediction}, lack fine details of swirls or shock waves, which explains the sharp decrease in accuracy for later time steps.
As the background medium is accurately computed, we can also validate that the model learns to link positions between different meshes well with the help of positional encoding and adding the position as a feature.
We also provide additional examples of jet predictions in the Appendix in \ref{sec:app_transformer-results}, where we continue to see worse results for faster jets.
For these, the turbulent jets make up a greater percentage of the sequences, showing that the computation of turbulent medium still shows room for improvement.

\section{Conclusion}

This work presents two distinct approaches for investigating SRMHD evolution with Neural Operators.
More specifically they aim at building surrogate models for the Black Hole Accretion Code (\texttt{BHAC}).

\paragraph{i)} A Physics-Informed Fourier Neural Operator (PINO) is trained to predict the evolution of \textit{resistive} SRMHD, with the Orszag-Tang vortex serving as the test case. Training is performed on Orszag-Tang vortices spanning resistivities in both the Sweet-Parker and fast reconnection regimes. Data supervision is sparse in the temporal domain, and the governing equations are embedded as an additional loss term imposed at finer temporal resolution. A clear improvement in model performance is observed in the sparsely supervised regime. Notably, the physics informed model successfully predicts plasmoid formation, whereas an otherwise identical data-only baseline fails to do so.

This approach suffers from several limitations: GPU memory pressure at high spatial resolutions, the stiffness of Ampère's law in the low-resistivity limit, and downsampling artifacts when training below the effective AMR resolution. Most critically, working in the strong-form formulation is in tension with the shock discontinuities present in the data.


\paragraph{ii)} An OFormer Neural Operator is trained to predict the evolution of Spine-Sheath Relativistic Jets on an AMR grid.
We were able to show that the model is successfully learning to natively encode different AMR input and output grids and can evolve SRMHD in time and space.
While continuous propagation from an initial step (rollout) and its convergence have not been tested, yet, the success of next-timestep predictions already shows the potential of such simulations.

However, our solution also suffers from large memory footprints and long training times which despite linear attention and Nyström sampling could present itself as a major hurdle when scaling the model in the future.
This holds especially true, if the lessons from the PINO are transferred to the OFormer AMR model which could be needed for better capturing finer details.

\vspace{1em}

\noindent Taken together, these two lines of investigation demonstrate complementary strengths of neural operator surrogates for relativistic MHD. Physics-informed training reduces the dependence on densely sampled simulation data and enables the recovery of physically meaningful structures that purely data-driven models miss. Operating natively on AMR grids, in turn, avoids the information loss of regridding and brings neural surrogates closer to the data representations already used by established solvers. A natural next step is to combine both directions -- integrating physics informed losses into the AMR-native framework -- to further improve the physical fidelity of the predictions.


\printbibliography

\appendix
\section*{Appendix}

\section{Details about the PINO for resistive SRMHD}

\subsection{Special-Relativistic Resistive MHD equations}
\label{sec:equations}

Maxwell's equations
\begin{gather}
    \nabla \cdot \mathbf{B} = 0 \label{eq:divB} \\
    \partial_t \mathbf{B} + \nabla \times \mathbf{E} = 0 \label{eq:FI} \\
    \nabla \cdot \mathbf{E} = q \label{eq:divE} \\
    -\partial_t \mathbf{E} + \nabla \times \mathbf{B} = \mathbf{J} \label{eq:Ampere}
\end{gather}
Ohm's law in the general inertial frame is given by
\begin{equation}
    \mathbf{J}= \frac{\gamma}{\eta} \ \left[\mathbf{E} + \mathbf{v} \times \mathbf{B} - (\mathbf{E}\cdot\mathbf{v})\,\mathbf{v} \right] + q\,\mathbf{v},
    \label{eq:Ohm}
\end{equation}
with $\gamma$ being the Lorentz factor and $\mathbf{v}$ the
velocity as measured by the inertial observer, and $\mathbf{u} = \gamma\,\mathbf{v}$
the spatial part of the 4-velocity.
The energy and momentum conservation laws are
\begin{equation}
    \partial_te+\nabla\cdot\mathbf{S}=0,
    \label{eq:em1}
\end{equation}
\begin{equation}
    \partial_t\mathbf{P}+\nabla\cdot\mathbf{\Pi}=0,
    \label{eq:em2}
\end{equation}
with the energy density 
\begin{equation}
    e = \frac{1}{2}(E^2+B^2)+w\gamma^2-p,
    \label{eq:energy_density}
\end{equation}
and the energy flux density $\mathbf{S}$ and and the momentum density $\mathbf{P}$
\begin{equation}
    \mathbf{S}=\mathbf{P}= \mathbf{E}\times\mathbf{B}+w\,\gamma^2\, \mathbf{v},
    \label{eq:energyfluxdensity}
\end{equation}
and the stress tensor
\begin{equation}
    \mathbf{\Pi}= -\mathbf{E}\mathbf{E}-\mathbf{B}\mathbf{B}+w\,\gamma^2\,\mathbf{v} \mathbf{v}+\left(\frac{1}{2}(E^2+B^2)+p\right)\,\mathbf{g}.
    \label{eq:momentumdensity}
\end{equation}
$\mathbf{g}$ is the
metric tensor of space.
The relativistic enthalpy per unit volume as measured in the rest frame of the fluid is
\begin{equation}
    w = \rho + \frac{\Gamma \, p}{\Gamma - 1},
    \label{eq:enthalpy}
\end{equation}
with $\Gamma$ being the adiabatic index.

The continuity equation is
\begin{equation}
    \partial_t\rho\,\gamma + \nabla\cdot(\rho\,\gamma\,\mathbf{v})=0.
    \label{eq:continuity}
\end{equation}
These equations close the system. \cite{Komissarov_2007}

\subsection{Orszag-Tang vortex}
\label{sec:app_OTV}

The Orszag-Tang vortex \cite{Orszag_Tang_1979} is a toy model for turbulent special-relativistic resisitve MHD flow.
It exhibits strong shocks as well as plasmoid formation.
The following section is based on \cite{Ripperda_2020}.

The used setup is a relativistic ideal gas with an adiabatic index of $\Gamma = \frac{4}{3}$, an initial uniform pressure $p = 10$ and initial rest mass density $\rho= 1$.

A $2.5$-dimensional grid $(x,y,z)$ is used.
The magnetic field $B=\nabla\times \mathbf{A}$ is initialized by a vector potential $\mathbf{A} = (0,\ 0,\ A_z)$ 
\begin{equation}
    A_z = \frac{1}{2}\cos{(2x)}+\cos{(y)}.
    \label{eq:OT_vec_pot}
\end{equation}
The initial velocity field $\mathbf{v} = (v_x,\ v_y,\ 0)$ is defined as
\begin{equation}
    \mathbf{v} = (-v_{max}\sin(y),\ v_{max}\sin (x),\ 0),
    \label{eq:OT_velocity}
\end{equation}
with $v_{max}={0.99\,c}/{\sqrt{2}}$. This ensures the maximum speed is limited by the speed of light.

The physical domain is set to $(x,\ y) \in\ [0,\ 2\pi]^2$ with periodic boundary conditions. The time domain is set to $t \in\ [0,\ 10]$, assuming natural units.

\subsection{FNO}
\label{sec:app_FNO}

The architecture of the Fourier Neural Operator is as follows (based on \cite{duruisseaux2026fourierneuraloperatorsexplained}). The input $a$ is lifted using a neural network $\mathcal{P}$. This produces a latent state $v_i$, where $i$ indexes the model layer. This is passed both through a linear transform $\mathcal{W}_i$ and the Fourier kernel. In the latter the state is Fourier transformed and a learnable linear transformation $R_i$ applied and the higher frequency modes truncated.
The results from both branches are added and an activation function $\sigma$ is multiplied.
\begin{equation}
    \mathcal{Q} \ \circ\ \sigma\,(\mathcal{W}_L+\mathcal{K}_L)\ \circ \ \dots \ \circ\ \sigma\,(\mathcal{W}_1+\mathcal{K}_1)\ \circ \mathcal{P},
    \label{eq:Fourier_structure}
\end{equation}
With the Fourier kernel operator $\mathcal{K}$
\begin{equation}
    (\mathcal{K}_i(\phi)v_i)(x) = \mathcal{F}^{-1}\big(R_i \cdot (\mathcal{F}v_i)\big)(x).
    \label{eq:Fourier_Kernel}
\end{equation}
In total the learnable entities of the FNO are $\mathcal{P}, \mathcal{Q}, \mathcal{W}_i, \mathcal{R}_i$.

FNO assumes periodicity in the in- and output, which conveniently is the case for the Orszag-Tang vortex.The temporal domain is padded to compensate for its inherent non-periodicity.

\subsection{PINO results}

\begin{figure}[H]
    \centering
    \includegraphics[width=\textwidth]{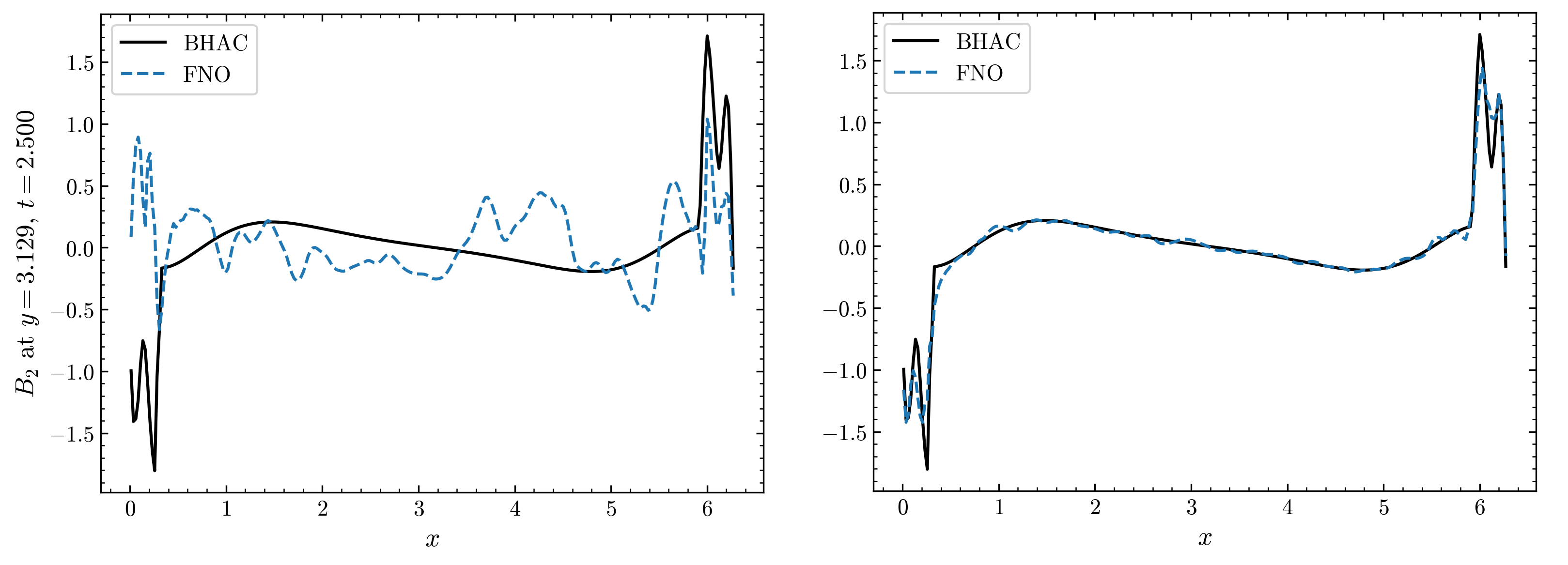}
    \caption{Slice of $B_y$ at $y = 3.129$ showing the model's performance at a timestep without data supervision, without (left) vs. with (right) PDE constraints.}
    \label{fig:B2_slice_vs}
\end{figure}

\begin{figure}[H]
    \centering
    \includegraphics[width=\textwidth]{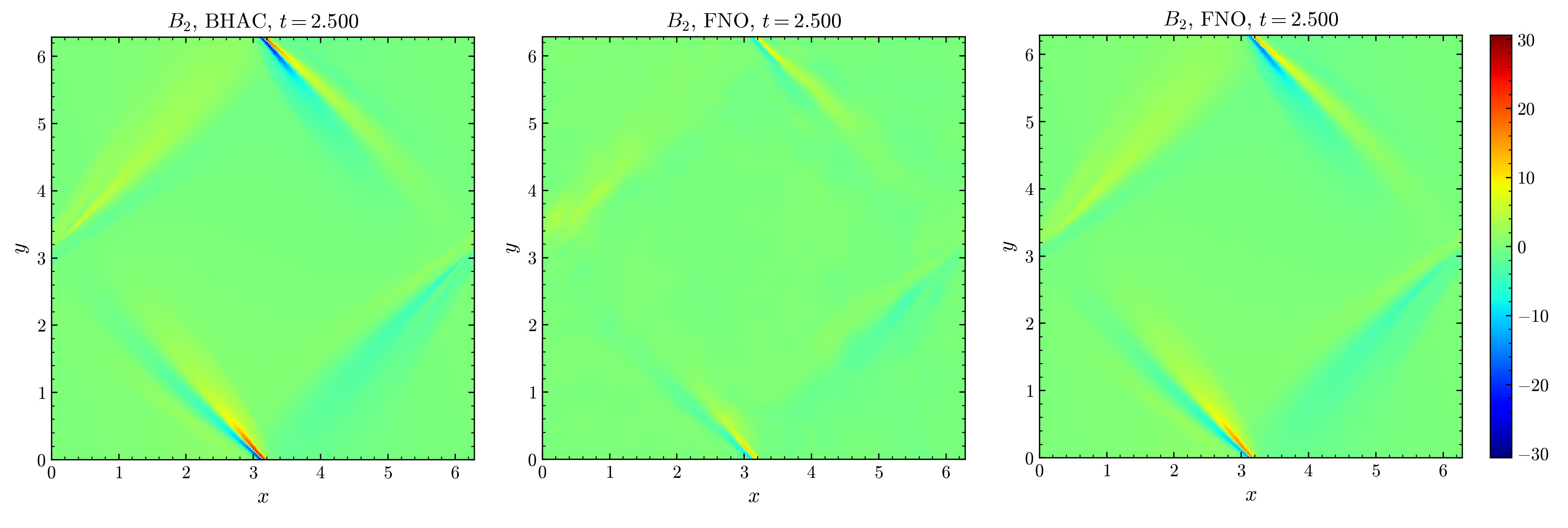}
    \caption{$B_y$ from \texttt{BHAC} (left) with the model's performance at a timestep without data supervision, without (middle) vs. with (right) PDE constraints.}
    \label{fig:By}
\end{figure}

\subsubsection{Plasmoid regime}

\begin{figure}[H]
    \centering
    \includegraphics[width=\textwidth]{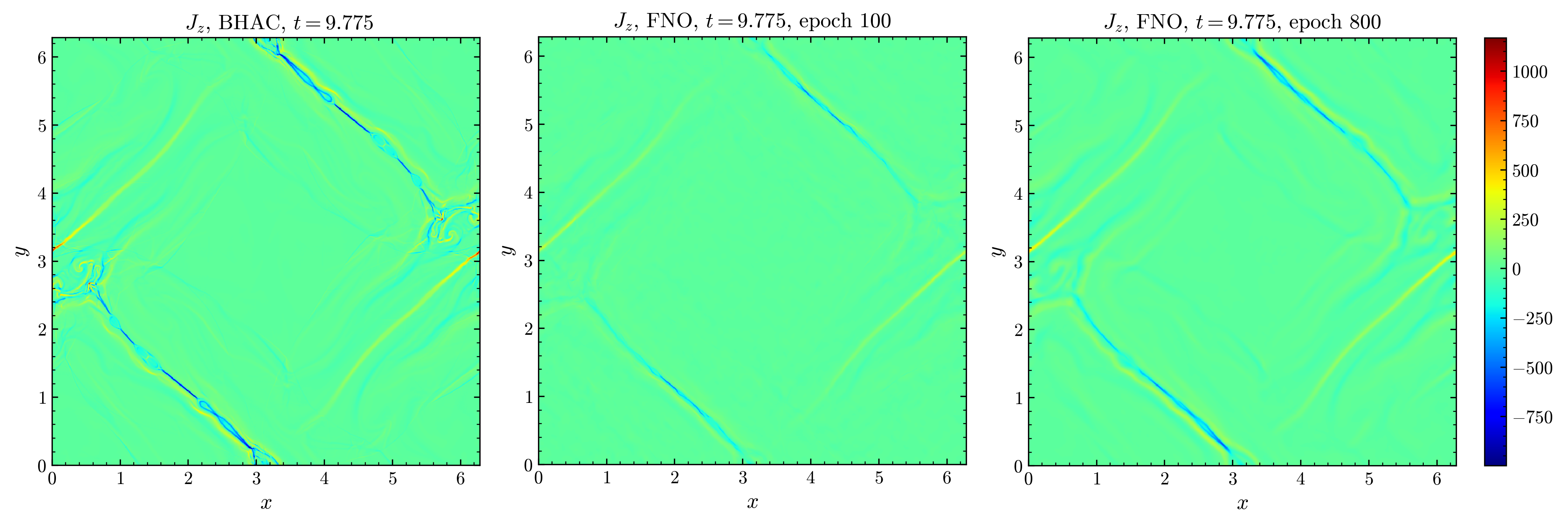}
    \caption{Electric current density $J_z$ at a timestep without data supervision, at resistivity $\eta=1.25\cdot10^{-4}$. Ground truth (left), the model's interpolation without PDE constraints (middle) and the model's predictions with PDE constraints enforced (right). Plasmoids are visible in the physics informed model, that were not present in the model without physics information.}
    \label{fig:Jz_00}
\end{figure}
\begin{figure}[H]
    \centering
    \includegraphics[width=\textwidth]{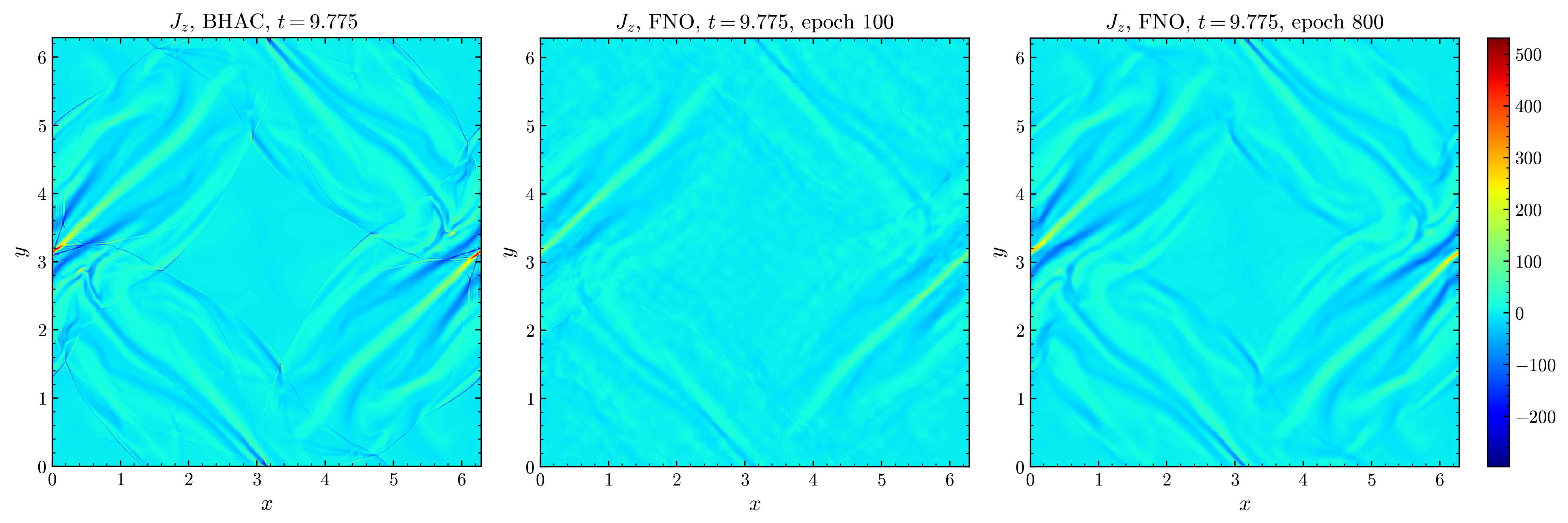}
    \caption{Electric current density $J_z$ at a timestep without data supervision, at resistivity $\eta=5.59\cdot10^{-4}$. Ground truth (left), the model's interpolation without PDE constraints (middle) and the model's predictions with PDE constraints enforced (right). A clear improvement in accuracy is seen from the physics informed model to the model without physics information.}
    \label{fig:Jz_03}
\end{figure}

\section{Details about the AMR-native Neural Operator}

\subsection{Spine-Sheath Relativistic Jet Dataset}
\label{sec:app_jets}

In order to investigate the propagation of relativistic jets with different initial conditions, i.e. injection velocity, density contrast with respect to the ambient medium and pressure ration, we build a large 2D library of propagating relativistic jets.
In the following, we provide details on the numerical setup.

\subsubsection{Numerical Model}

The simulations presented in this work are performed using \texttt{BHAC} \cite{Porth2017, Olivares2019, Ripperda2019}, solving the equations of ideal special relativistic magnetohydrodynamics (SRMHD) using adaptive mesh refinement and cylindrical geometry, appropriate to describe the transverse structure of relativistic jets.
This drop in physical resistivity leads to a change in the MHD equations presented in \autoref{sec:equations}.
\autoref{eq:Ohm} reduces to
\begin{equation}
    \mathbf{E}+ \mathbf{v}\times \mathbf{B} =0.
\end{equation}
This removes the dependence of $\mathbf{J}$ from $\mathbf{E}$ in \autoref{eq:Ampere} and $\mathbf{E}$ stops being an independent dynamical variable.

The numerical grid covers a range of $100\times200$ jet radii, $R_j$ in radial and z-direction.
We make full use of the AMR capabilities of \texttt{BHAC} code and use a base grid of $256\times256$ pixels with four additional levels of AMR.
This leads to an effective resolution of $4096\times4096$ pixels corresponding to a solution of $\sim40$ cells per jet radii in radial and $\sim20$ cells per jet radii in z-direction. 

\paragraph{Jet Injection Conditions:}
At the jet inlet, we prescribe radial profiles based on a core-envelope configuration of \cite{Komissarov_2007,komissarov2015stationary,fromm2017radiative}.
The initial state is defined by the gas pressure $p(r)$ and the co-moving azimuthal magnetic field $b_\phi = B_\phi / \Gamma$.
The pressure profile is specified as
\begin{equation}
p(r) =
\begin{cases}
d_k p_0 \left[ \alpha + \frac{2}{\beta_m} \left( 1 - (r/r_m)^2 \right) \right],
& r < r_m, \\
\alpha d_k p_0,
& r_m < r < r_j, \\
p_0,
& r > r_j ,
\end{cases}
\end{equation}
where $r_m$ denotes the radius of the jet core and $r_j$ the total jet-radius and $d_k$ the pressure mismatch between the jet and the ambient medium.
The azimuthal magnetic field follows
\begin{equation}
b_\phi(r) =
\begin{cases}
b_m (r/r_m),
& r < r_m, \\
b_m (r_m/r),
& r_m < r < r_j, \\
0,
& r > r_j ,
\end{cases}
\end{equation}
ensuring a continuous magnetic structure across the core-sheath transition.
Here, we define
\begin{equation}
\beta_m = \frac{2 p_0}{b_m^2},
\qquad
\alpha = 1 - \frac{1}{\beta_m}
\left( \frac{r_m}{r_j} \right)^2 ,
\end{equation}

\subsubsection{Velocity and Density}

The magnetic and pressure profiles described above can be combined with arbitrary velocity and density distributions.
In this work, we adopt a radial profile for the bulk Lorentz factor of the form
\begin{equation}
\Gamma(r) = \Gamma_0 \left[ 1 - (r/r_j)^\mu \right] + (r/r_j)^\mu ,
\end{equation}
where $\Gamma_0$ denotes the Lorentz factor on the jet axis.
For sufficiently large values of $\mu$ (specifically $\mu > 8$), this prescription produces an almost uniform Lorentz factor within the jet, while the surrounding medium remains stationary.
The mass density is initialized according to
\begin{equation}
\rho(r) =
\begin{cases}
\eta \rho_a,
& r < r_j, \\
\rho_a,
& r > r_j ,
\end{cases}
\end{equation}
where $\rho_a$ is the ambient density at the jet nozzle and $\eta$ defines the jet-to-ambient density contrast.
The ambient medium is assumed to be constant in density and pressure.
If not mentioned otherwise, the ambient density is set to $\rho_a=1$ and the ambient pressure to $p_0$.
We close our system of equations using an ideal equation of state with an adiabatic index of $\hat{\gamma}=4/3$.
In addition to the adiabatic index we fix the jet's outer radius to $r_j=1\,R_j$ and the jet's core to $r_m=0.27\,R_j$.
Furthermore, to ensure the same ambient pressure for all models we use fixed values for $\beta_m=0.6$ and $\beta_m=0.138$ which leads to an ambient pressure of $p_0=0.02484$ given in code units.

The open parameters for our model library are therefore the jet velocity, the jet density ratio, $\eta$, and the over-pressure parameter, $d_k$.
We sample the velocity and density ratio logarithmically over eight grid points, while for the over-pressure we choose seven points on a linear scale.
The boundaries are $0.5\leq v\leq 0.995$, $-3\leq\eta\leq-0.3$, and $1\leq d_k\leq 4$.
In total, our library consists of 448 high-resolution simulations of propagating relativistic jets.
During the run of each simulation we write out a snapshot every 10 time units while the total run time for each model is 500 time units.
Therefore, the resulting data set consists of $22400$ SRMHD snapshot pairs.

\subsection{\texttt{BHAC} Fields for Relativistic Jets}
\label{sec:app_jet-vars}

\subsubsection*{Primitive Variables}

Primitive variables are the physical variables included in \texttt{BHAC} snapshots that govern the current state of the simulation, i.e., no other variables are needed to continue the simulation except for potentially global parameters.

\begin{table}[!ht]
    \centering
    \caption{Principle measures in \texttt{BHAC} snapshots and presence in data set.}
    \begin{tabular}{|c|c|c|c|}
    \hline
        \textbf{Variable Name} & \textbf{Symbol} & \textbf{Included?} & \textbf{Encoded?} \\\hline
        Gas Pressure & $p$ & yes & yes, $\log(p)$ \\\hline
        Radial Spatial Velocity & $u_1 = \gamma \cdot v_1$ & yes & no \\\hline
        Spatial Velocity in $z$ & $u_2 = \gamma \cdot v_2$ & yes & no \\\hline
        Spatial Velocity in $\phi$ & $u_3 = \gamma \cdot v_3$ & yes & no \\\hline
        Azimuthal Magnetic Field in $r$ & $b_1$ & no, constantly $0$ & no \\\hline
        Azimuthal Magnetic Field in $z$ & $b_2$ & no, constantly $0$ & no \\\hline
        Azimuthal Magnetic Field in $\phi$ & $b_3$ & yes & no \\\hline
        Ambient pressure & $\rho$ & yes & yes, $\log(\rho)$ \\\hline
    \end{tabular}
\end{table}

\subsubsection*{Auxiliary variables}

Auxiliary variables are either helper variables or are computable from primitive and global parameters and are therefore not directly needed for continuing the simulation.
Currently, none of these variables are used in the simulator.

\begin{table}[!ht]
    \centering
    \caption{Auxiliary measures in \texttt{BHAC} snapshots and presence in data set.}
    \begin{tabular}{|c|c|c|c|}
    \hline
        \textbf{Variable Name} & \textbf{Symbol} & \textbf{Included?} & \textbf{Encoded?} \\\hline
        Fluid Entropy & $s$ & no & no \\\hline
        Tracer Variable of Jet & $\text{tr}1$ & no & no \\\hline
        Lorentz factor & $\Gamma$ & no & no \\\hline
        Relativistic Enthalpy & $\xi$ & no & no \\\hline
        Magnetic Energy Density & $B^2$ & no & no \\\hline
        Magnetic Field Divergence & $\vec{\nabla}\vec{b}$ & no & no \\\hline
    \end{tabular}
\end{table}

\subsubsection*{Global Parameters}

Global parameters define the simulation environment and the initial conditions of the jet.
Currently, only changing parameters are included in the simulation as constants can be learned by the simulating model.

\begin{table}[!ht]
    \centering
    \caption{Global parameters in \texttt{BHAC} snapshots present in data set.}
    \begin{tabular}{|c|c|c|c|}
    \hline
        \textbf{Variable Name} & \textbf{Symbol}  \\\hline
        Simulation Time & $t$ \\\hline
        Fluid injection velocity & $v_b$ \\\hline
        Jet density ratio & $\eta$ \\\hline
        Pressure mismatch jet/ambient medium & $d_k$ \\\hline
    \end{tabular}
\end{table}

\subsection{Model Architecture}

\label{sec:app_oformer-architecture}
\begin{figure}[H]
    \centering
    \includegraphics[width=0.8\linewidth]{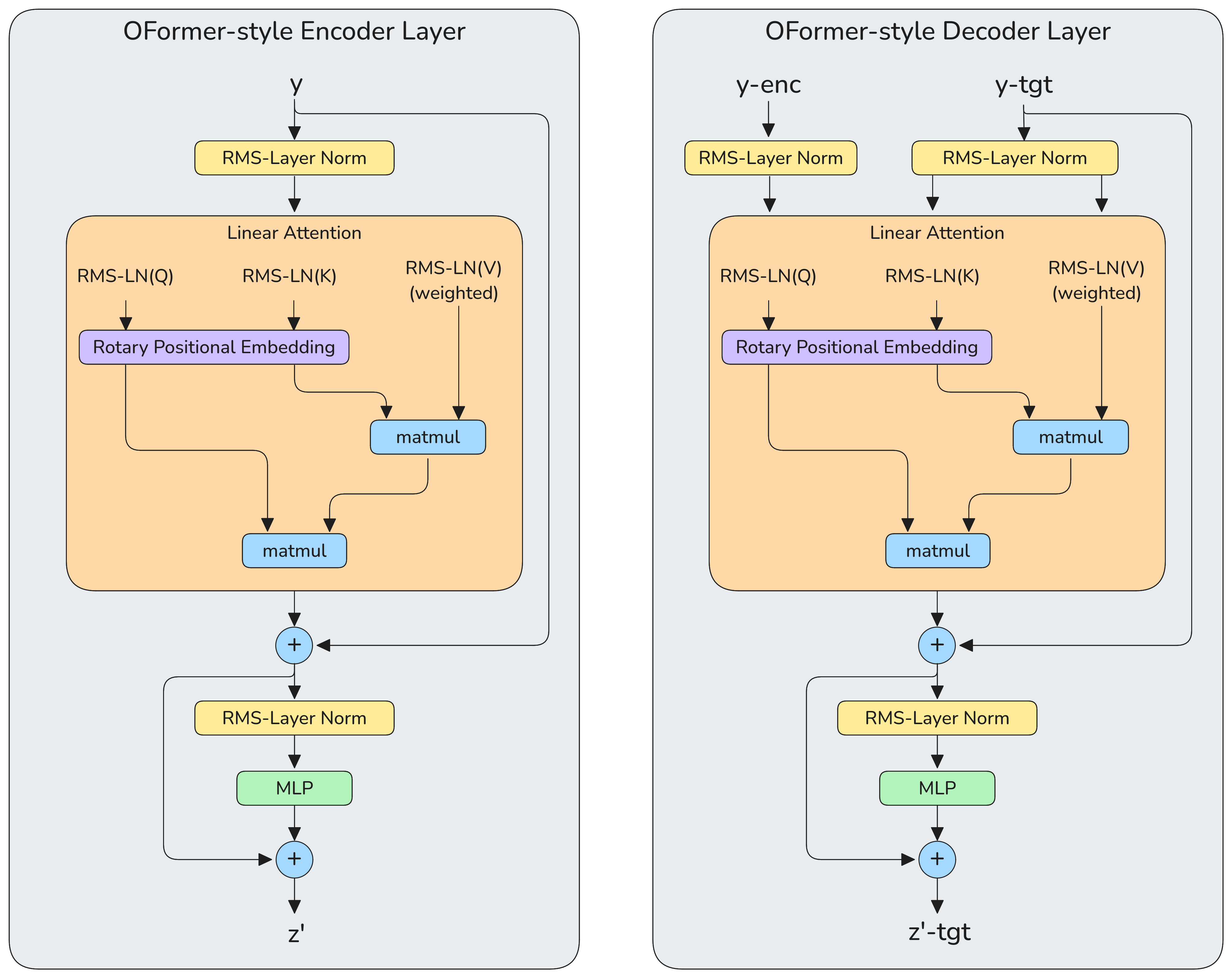}
    \caption{Depiction of the Encoder layer (left) and Decoder layer (right) of the OFormer architecture. The overall architecture stays close to vanilla transformer layers with pre-normalization, but replaces the attention with a Galerkin-style linear attention.}
    \label{fig:oformer-layers}
\end{figure}

\subsection{Identity baseline and overfitted OFormer}
\label{sec:app_oformer-baselines}

As the problem requires an AMR grid, calculating the identity function is not trivially possible.
However, the AMR grid in this instance does not change drastically, thus we find the common points between the input and output and calculate the loss between the input and output queried at these points.

The overfitted model was trained for $6000$ epochs on one simulation which is equivalent to $30000$ steps.
All other parameters are chosen to be equal to the non-overfitted model.
The last model was then evaluated on the same data set it was trained on, producing the overfitted OFormer baseline.

\pagebreak
\subsection{Additional Prediction Results}
\label{sec:app_transformer-results}

\begin{figure}[H]
    \centering
    \includegraphics[width=0.9\linewidth]{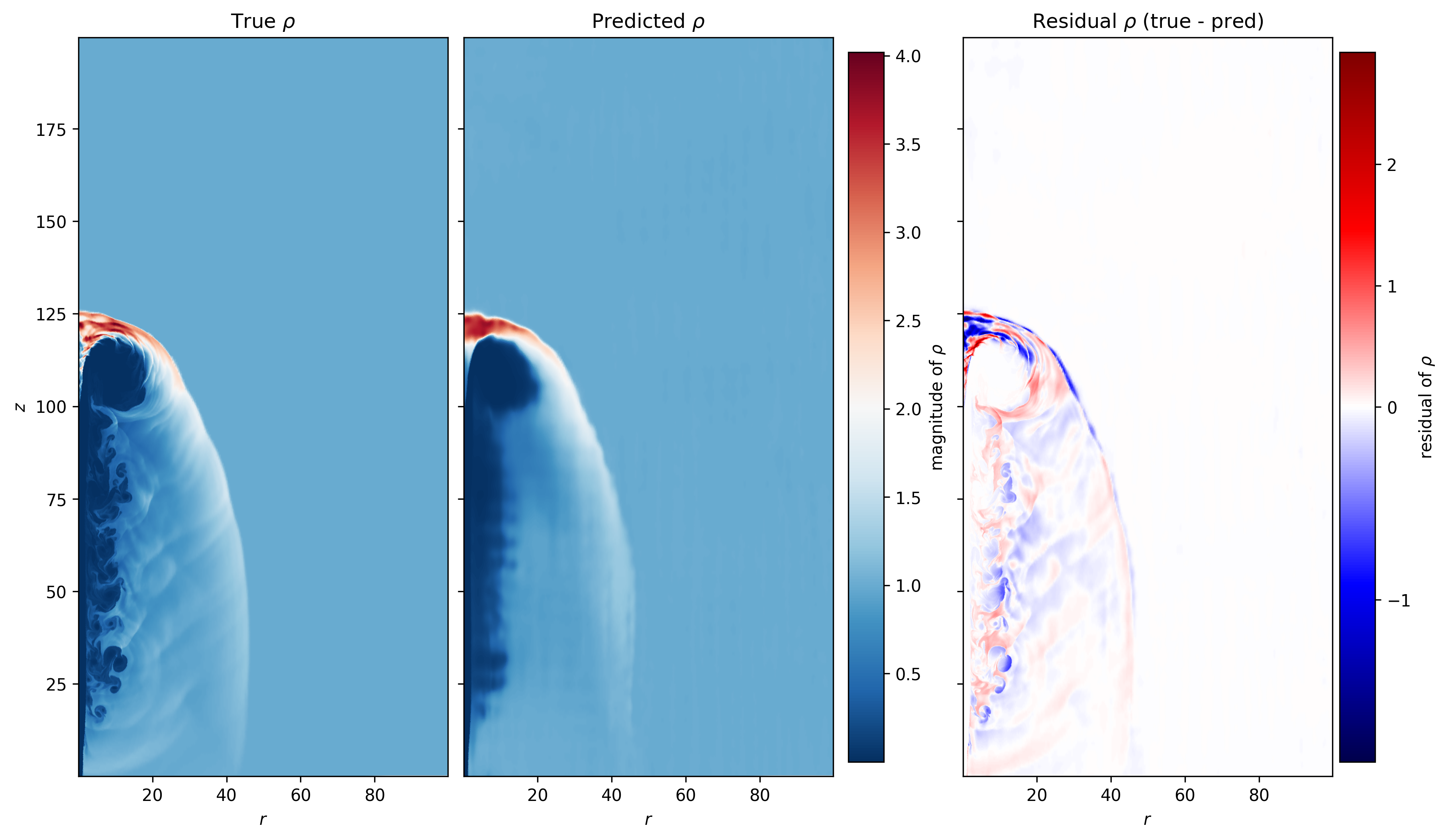}
    \caption{Prediction of $\rho$ at $t=20$. The parameters for the jet are $d_k=4$, $v_b=0.995$, $\eta=0.006$. ($\text{L}_2=0.303$, $\text{L}_\infty=0.680$, $\text{L}_{2, \rho}=0.279$)}
\end{figure}

\begin{figure}[H]
    \centering
    \includegraphics[width=0.9\linewidth]{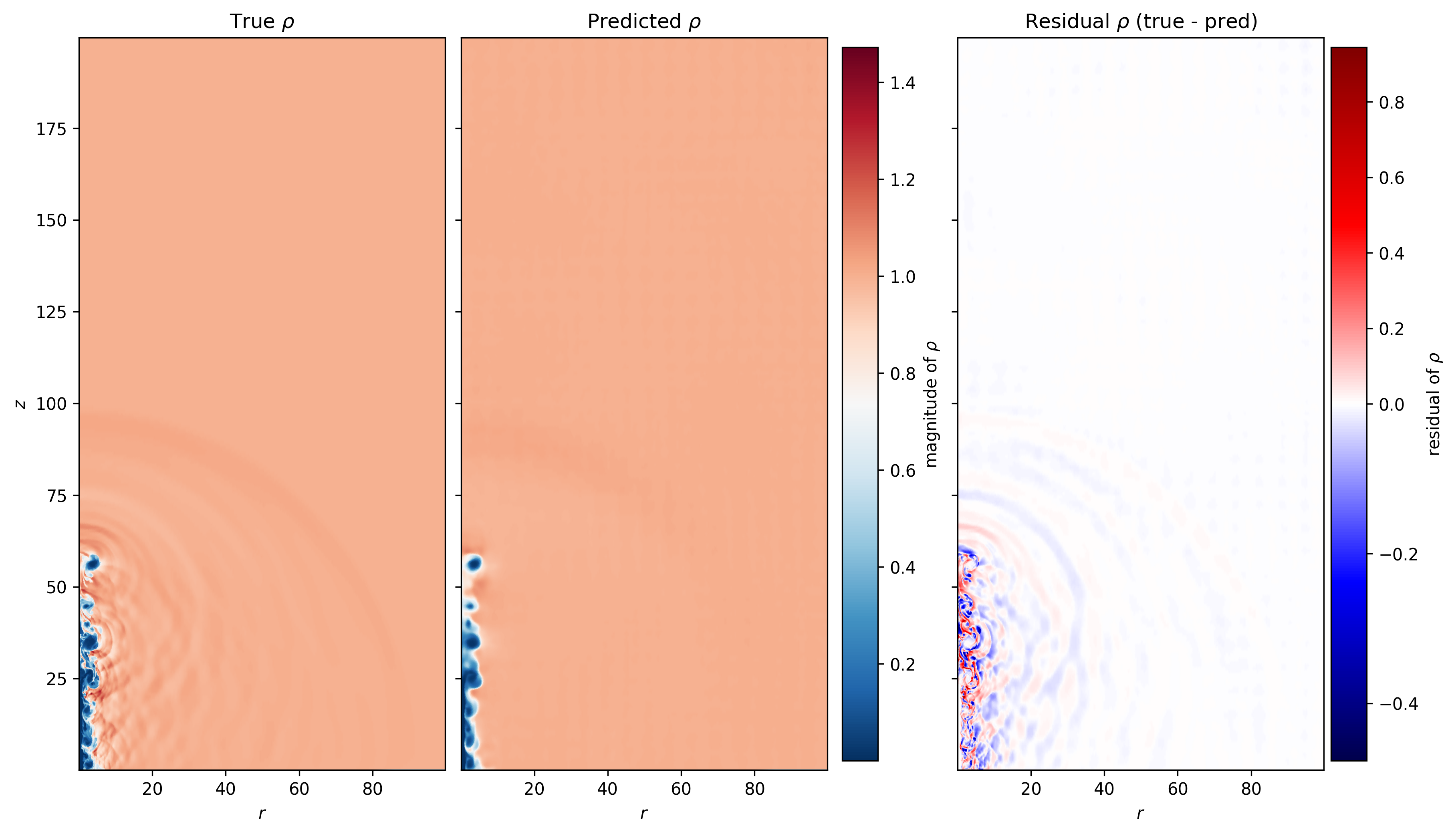}
    \caption{Prediction of $\rho$ at $t=50$. The parameters for the jet are $d_k=2$, $v_b=0.672$, $\eta=0.002$. ($\text{L}_2=0.395$, $\text{L}_\infty=0.722$, $\text{L}_{2, \rho}=0.2540$)}
\end{figure}

\begin{figure}[H]
    \centering
    \includegraphics[width=0.9\linewidth]{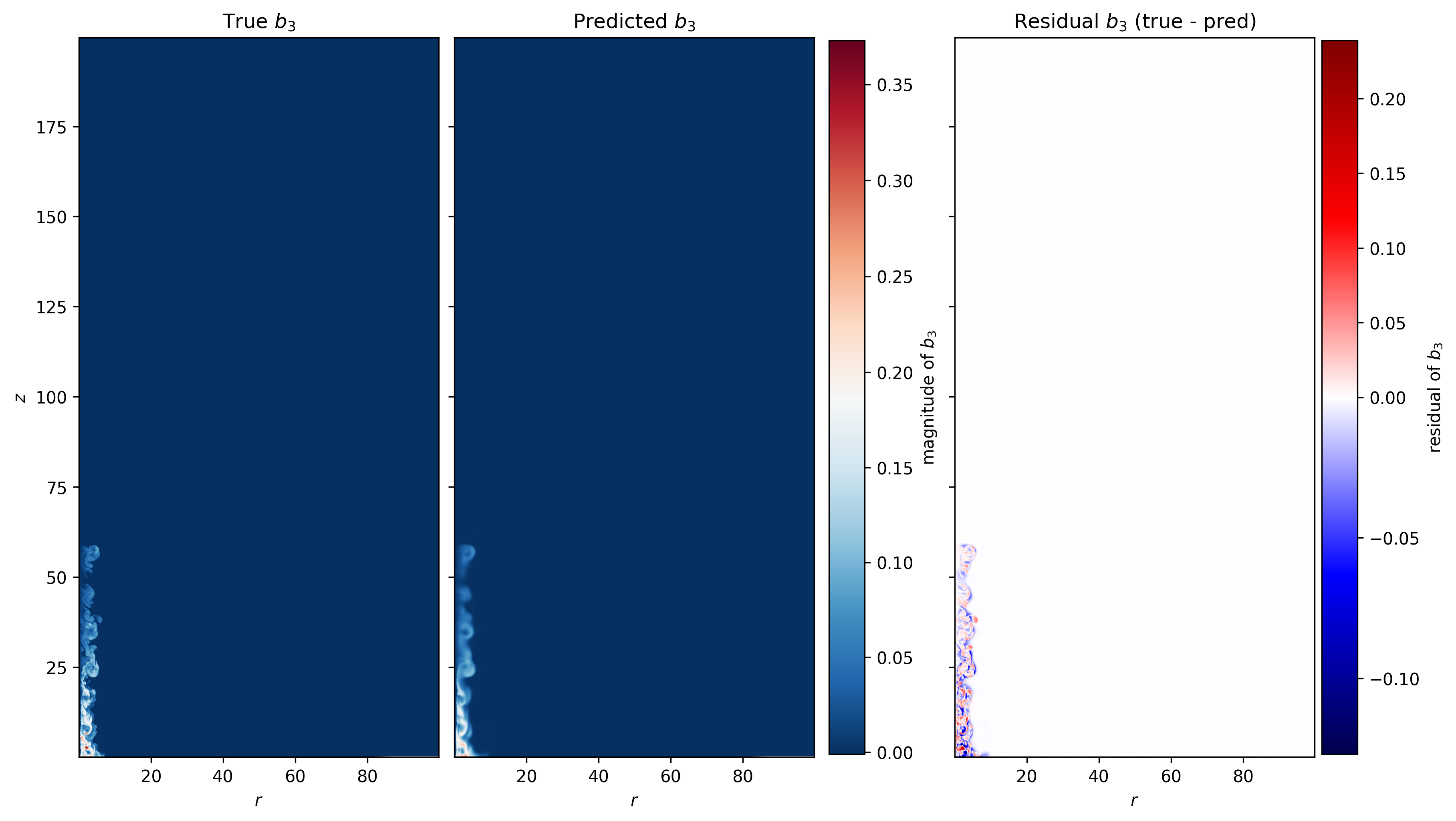}
    \caption{Prediction of $b_3$ at $t=50$. The parameters for the jet are $d_k=2$, $v_b=0.672$, $\eta=0.002$. ($\text{L}_2=0.395$, $\text{L}_\infty=0.722$, $\text{L}_{2, b_3}=0.3446$)}
\end{figure}

\begin{figure}[H]
    \centering
    \includegraphics[width=0.9\linewidth]{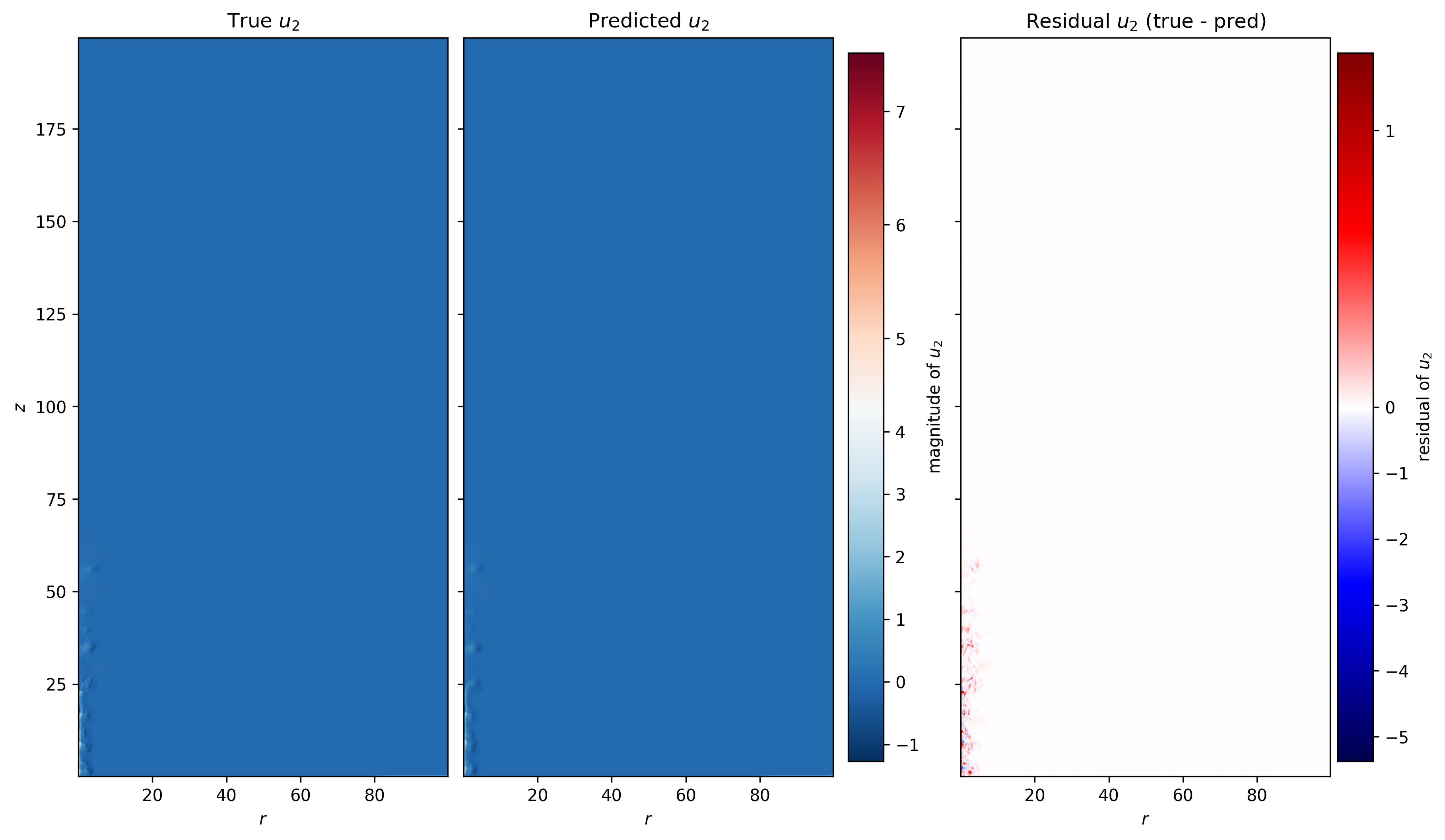}
    \caption{Prediction of $u_2$ at $t=50$. The parameters for the jet are $d_k=2$, $v_b=0.672$, $\eta=0.002$. ($\text{L}_2=0.395$, $\text{L}_\infty=0.722$, $\text{L}_{2, u_2}=0.2091$)}
\end{figure}

\end{document}